\documentclass[useAMS,usenatbib,fleqn]{mn2e}

\usepackage{graphicx,amssymb}
\usepackage{epsf,psfig}
\usepackage{natbib}
\usepackage{amsmath}

\title[Dynamics of a tidally limited star cluster]
{Revealing the escape mechanism of three-dimensional orbits in a tidally limited star cluster}

\author[Euaggelos E. Zotos]{Euaggelos E. Zotos\thanks{E-mail: evzotos@physics.auth.gr} \\
Department of Physics, School of Science, Aristotle University of Thessaloniki,
GR-541 24, Thessaloniki, Greece}

\begin{document}

\date{Accepted 2104 October 9. Received 2014 October 8; in original form 2014 July 24}

\pubyear{2015} \volume{446} \pagerange{770--792} 

\setcounter{page}{770}

\maketitle

\label{firstpage}

\begin{abstract}
The aim of this work is to explore the escape process of three-dimensional orbits in a star cluster rotating around its parent galaxy in a circular orbit. The gravitational field of the cluster is represented by a smooth, spherically symmetric Plummer potential, while the tidal approximation was used to model the steady tidal field of the galaxy. We conduct a thorough numerical analysis distinguishing between regular and chaotic orbits as well as between trapped and escaping orbits, considering only unbounded motion for several energy levels. It is of particular interest to locate the escape basins towards the two exit channels and relate them with the corresponding escape times of the orbits. For this purpose, we split our investigation into three cases depending on the initial value of the $z$ coordinate which was used for launching the stars. The most noticeable finding is that the majority of stars initiated very close to the primary $(x,y)$ plane move in chaotic orbits and they remain trapped for vast time intervals, while orbits with relatively high values of $z_0$ on the other hand, form well-defined basins of escape. It was also observed, that for energy levels close to the critical escape energy the escape rates of orbits are large, while for much higher values of energy most of the orbits have low escape periods or they escape immediately to infinity. We hope our outcomes to be useful for a further understanding of the dissolution process and the escape mechanism in open star clusters.
\end{abstract}

\begin{keywords}
stars: kinematics and dynamics –- galaxies: star clusters.
\end{keywords}

\section{Introduction}
\label{intro}

Star clusters are formed in various regions of a galaxy and they have the tendency to gradually lose mass, mainly due to dynamical interactions, until they completely dissolved. A considerable proportion of stars that are born in stellar clusters become part of the disk and halo populations of the Milky Way as the clusters dissolve over time. The dissolution process of star clusters is a rather old field of stellar dynamics which however, still remains very active and challenging. Theoretical aspects on this intricate subject like the evaporation process about the movement of stars above the escape velocity and the time-scale of relaxation which determines the rate of the dynamical evolution of a star cluster were the first that studied analytically \citep[e.g.,][]{A38,S40}. Later on, the properties of stars that have been scattered above the critical escape energy were examined \citep{K59}, while a year later the importance of close encounters between stars for the rate of mass loss of a star cluster was pointed out by \citet{H60}.

As time went by the computational capabilities grew exponentially, while at the same time the numerical codes were evolved and also refined thus allowing us to use $N$-body simulations in order to obtain a more complete and accurate view regarding the formation as well as the evolution of star clusters \citep[e.g.,][]{KTG93,MH97,TAP97,A99,HPA05,GLH13}. At this point we should emphasize, that all the above-mentioned references on star cluster $N$-body simulations are exemplary rather than exhaustive. It is worth mentioning that one of the most famous and sophisticated collisional $N$-body codes that takes into account many important dynamical features about the structure of the star cluster is the NBODY6 \citep{A03}.

It is widely believed that the vast majority of stars of a galaxy are born in star clusters \citep{LL03}, while recent studies \citep[e.g.,][]{BBG10,K12} shed some new light on the procedure of how stars are actually born and form stellar clusters. Open stellar clusters are formed by gravitationally bounded stars however, all clusters dissolve over time because of two main reasons: (i) the two-body process that dynamically affects all stars of a cluster causing them to obtain escape velocities and (ii) the strong tidal forces due to the external potential of the parent galaxy in which clusters are embedded. In particular, the tidal field of the parent galaxy can significantly influence the dynamical behaviour of the cluster itself \citep{RMH97}.

Stars escape from tidally limited star clusters following a two stages process. During first stage, stars are being scattered into the escaping phase space by two-body encounters, while at second stage they escape through the channels in the open equipotential surface. The time required for a star to complete the first stage strongly depends on the relaxation time, while on the other hand, the time needed for a star to complete stage two is mainly related with the energy of the star. \citet{FH00} derived an expression for the escape times of stars in second stage which have just finished stage one, while \citet{B01} exploited these results in order to address the important issue of dissolution time of star clusters moving in circular orbits around their parent galaxy. Moreover, other interesting issues have also been studied like the velocity of escaping stars \citep[e.g.,][]{KG07,SN11,LLZ12}, or even their population type \citep{R97}.

When stars escape from a star cluster they are captured by the gravitational field of the parent galaxy and they usually form extended complicated formations called tidal tails or tidal arms. These special stellar structures have been observed in the Milky Way \citep[e.g.,][]{GFI95,KSL97,LS97,OGR01,ROG02,LLF03,BEI06} and also have been modelled and investigated \citep[e.g.,][]{JSH99,YL02,DOG04,KGO04,CMM05,dMCM05,LLS06,CWK07,FEB07,MCd07,KMH08,JBPE09,KKBH10}.

Escaping particles from dynamical systems is a subject to which has been devoted many studies over the years. Especially the issue of escapes in Hamiltonian systems is directly related to the problem of chaotic scattering which has been an active field of research over the last decades and it still remains open \citep[e.g.,][]{BGOB88,JS88,CK92,BST98,ML02,SASL06,SSL07,SS08,SHSL09,SS10}. The problem of escape is a classical problem in simple Hamiltonian nonlinear systems \citep[e.g.,][]{AVS01,AS03,AVS09,BBS08,BSBS12,Z14} as well as in dynamical astronomy \citep[e.g.,][]{HB83,BTS96,BST98,dML00,N04,N05,Z12}.

The chaotic dynamics of a star cluster embedded in the tidal field of a parent galaxy was investigated in \citet{EJSP08} (hereafter Paper I). Conducting a thorough scanning of the available phase space the authors managed to obtain the basins of escape and the respective escape rates of the orbits, revealing that the higher escape times correspond to initial conditions of orbits near the fractal basin boundaries. In the present work we use the analysis of the 2D system presented in Paper I as a starting point and we expand our exploration into three dimensions in an attempt to reveal the escape mechanism of three-dimensional orbits in a tidally limited star cluster.

The present article is organized as follows: In Section \ref{mod} we present in detail the structure and the properties of the tidal approximation model. All the computational methods we used in order to determine the character of orbits are described in Section \ref{cometh}. In the following Section, we conduct a thorough numerical investigation revealing the bounded and escaping regions of the cluster and how they are affected by the total orbital energy (Jacobi integral), as well as by the initial value of the $z$-coordinate of the three-dimensional orbits. In Section \ref{tc} we try to theoretically explain and justify the observed phenomenon of trapped chaos. Our paper ends with Section \ref{disc}, where the discussion and the conclusions of this research are given.

\section{The tidal approximation model}
\label{mod}

According to the ``tidal approximation" theory, we assume that the star cluster moves on a circular orbit around the galactic center at constant angular velocity $\omega$. This assumption allows us to apply the epicyclic approximation in order to compute linear tidal forces acting on all stars of the star cluster. In this case, the most appropriate system of coordinates is a rotating, accelerated reference frame \citep{C42} in which both the galactic center and the star cluster are at rest, while the origin of the coordinates is at the center of the star cluster which is located at the minimum of the effective potential of the galaxy, so that the position of the galactic center is $\left(-R_{\rm g},0,0\right)$, where $R_{\rm g}$ is the radius of the circular orbit. If the star cluster follows a more general orbit (i.e., elliptical) around its parent galaxy, then no integral of motion comparable to the Jacobi integral is known and consequently all the numerical calculations become more difficult to be performed.

For the description of the properties of the spherically symmetric star cluster, we use the simple, analytical and self-consistent Plummer potential
\begin{equation}
\Phi_{\rm cl}(x,y,z) = \frac{- G M_{\rm cl}}{\sqrt{x^2 + y^2 + z^2 + r_{\rm Pl}^2}},
\label{potcl}
\end{equation}
where $G$ is the gravitational constant, while $M_{\rm cl}$ is the total mass of the star cluster. The corresponding Plummer radius is $r_{\rm Pl} = 0.182$ (for more details about the isolated Plummer potential see the Appendix A in Paper I). The Plummer radius was chosen in such a way that model (\ref{potcl}) to be the best fit to a $W_0 = 4$ King model, which fills completely the Roche lobe in the tidal field.

Taking into account the relationships connecting the tidal forces with the epicyclic frequency $\kappa$ and the vertical frequency $\nu$ \citep[see for more details,][]{BT08} the basic equations of motion are
\begin{align}
\ddot{x} &= - \frac{\partial \Phi_{\rm cl}}{\partial x} - \left(\kappa^2 - 4 \omega^2 \right)x + 2 \omega \dot{y},\\
\ddot{y} &= - \frac{\partial \Phi_{\rm cl}}{\partial y} - 2 \omega \dot{x},\\
\ddot{z} &= - \frac{\partial \Phi_{\rm cl}}{\partial z} - \nu^2 z,
\label{eqmot}
\end{align}
where, as usual, the dot indicates derivative with respect to the time. We observe that both centrifugal and Coriolis forces appear in the above equations of motion due to the fact that the coordinate system is rotating. The total effective potential corresponding to the equations of motion is
\begin{equation}
\Phi_{\rm eff}(x,y,z) = \Phi_{\rm cl}(x,y,z) + \frac{1}{2}\left(\kappa^2 - 4\omega^2 \right) x^2 + \frac{1}{2}\nu^2 z^2.
\label{eff}
\end{equation}
The effective potential (\ref{eff}) has two Lagrangian saddle points $L_1$ and $L_2$ which are located at $(x,y,z) = (-r_{\rm t}, 0, 0)$ and $(x,y,z) = (r_{\rm t}, 0, 0)$, respectively where
\begin{equation}
r_{\rm t} = \left(\frac{G M_{\rm cl}}{4\omega^2 - \kappa^2}\right)^{1/3},
\label{rt}
\end{equation}
is the tidal radius \citep{K62} which provides a fundamental scale length of the star cluster.

\begin{figure}
\includegraphics[width=\hsize]{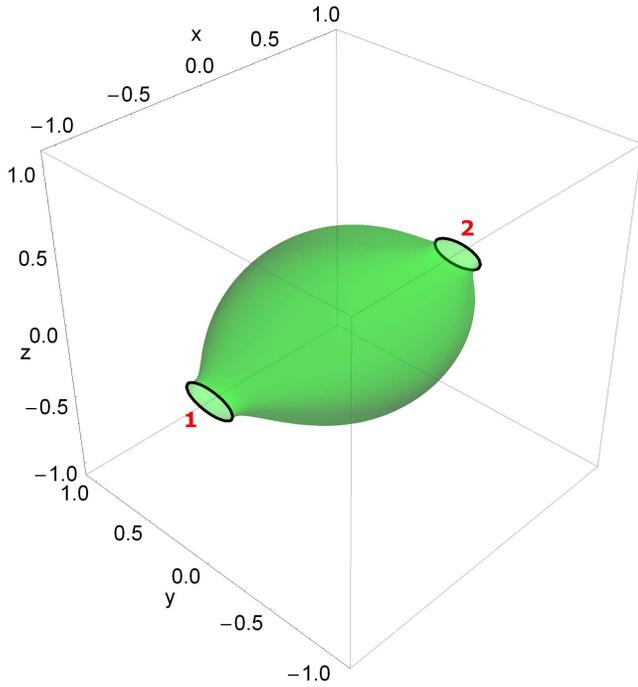}
\caption{The three-dimensional equipotential surface of the effective potential $\Phi_{\rm eff}(x,y,z) = E$ when $\widehat{C} = 0.01$. The escaping orbits leak out through the exit channels 1 and 2 of the equipotential surface passing either $L_1$ or $L_2$ Lagrangian points, respectively.}
\label{surf3d}
\end{figure}

\begin{figure*}
\centering
\resizebox{0.9\hsize}{!}{\includegraphics{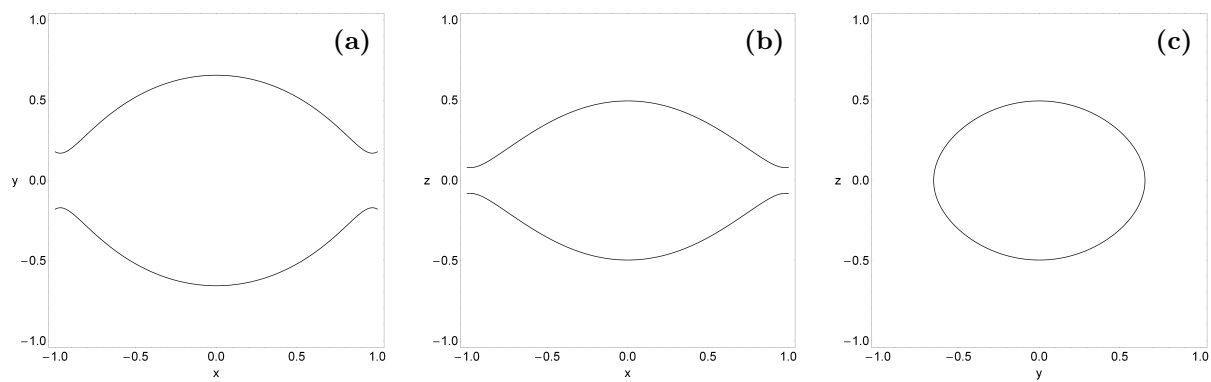}}
\caption{Contours of the projections of the equipotential surface $\Phi_{\rm eff}(x,y,z) = E$ on the three primary planes.}
\label{isop}
\end{figure*}

Furthermore, the equations of motion admit the isolating integral of motion
\begin{equation}
C = \frac{1}{2} \left(\dot{x}^2 + \dot{y}^2 + \dot{z}^2 \right) + \Phi_{\rm eff}(x,y,z) = E,
\label{ham}
\end{equation}
where $\dot{x}$, $\dot{y}$ and $\dot{z}$ are the momenta per unit mass, conjugate to $x$, $y$ and $z$, respectively, while $E$ is the numerical value of the Jacobian integral, which is conserved. The numerical value of the effective potential at the two Lagrangian points $\Phi_{\rm eff}(-1,0,0)$ and $\Phi_{\rm eff}(1,0,0)$ yields to a critical Jacobi constant $C_L = -3.264444506473746$, which can be used to define a dimensionless energy parameter as
\begin{equation}
\widehat{C} = \frac{C_L - E}{C_L},
\label{chat}
\end{equation}
where $E < 0$ is some other value of the Jacobian. The dimensionless energy parameter $\widehat{C}$ makes the reference to energy levels more convenient. For $E = C_L$ the equipotential surface encloses the critical volume, while for a Jacobian value $E > C_L$, or in other words $\widehat{C} > 0$, the equipotential surface is open and consequently stars can escape from the cluster. In Fig. \ref{surf3d} we present a plot of the three-dimensional equipotential surface $\Phi_{\rm eff}(x,y,z) = E$ when $\widehat{C} = 0.01$. We observe, the two openings (exit channels or throats) through which the particles can leak out. In fact, we may say that these two exits act as hoses connecting the interior of the cluster with the ``outside world". Exit channel 1 (negative $x$-direction) indicates escape towards the galactic center, while channel 2 (positive $x$-direction) indicates escape towards infinity. The extent of the equipotential surface in the positive $z$ direction is roughly 0.5 pc, while it is 1 pc in the positive $x$ direction and $2/3$ pc in the positive $y$ direction. Moreover, Fig. \ref{isop}(a-c) shows the projections of the equipotential surface $\Phi_{\rm eff}(x,y,z) = E$ on the three primary planes $(x,y)$, $(x,z)$ and $(y,z)$, respectively.

We use the dimensionless system of units introduced in Paper I according to which $G = 1$, $\omega = 1$ and $M_{\rm cl} = 4 \omega^2 - \kappa^2 = 2.2$. In addition, the unit of length is one tidal radius $r_{\rm t} = 1$ pc, while the time unit is equal to 1 Myr. In physical units we have $G = 1/222.3$ pc$^3$/M$_{\odot}$/Myr$^2$ therefore, we may set the mass of the cluster and the tidal radius to realistic values to obtain the scaling factors for mass and length. In particular, to pin down a specific model, we could assume the galactocentric distance and a model for the potential of the galaxy, which will allow us to find the value of $\omega$. If also we assume the mass of the cluster, then this immediately gives us the conversion of mass units. The characteristic frequencies $\omega$, $\kappa$ and $\nu$ on the other hand, are related to the galactic gravitational potential and in the solar neighborhood can be expressed as functions of the Oort's constants \citep[see e.g.,][]{BT08} as follows
\begin{align}
\omega^2 &= \left(A - B\right)^2, \\
\kappa^2 &= - 4B\left(A - B\right)^2, \\
\nu^2 &= 4\pi G \rho_{\rm g} + 2 \left(A^2 - B^2\right),
\label{freqs}
\end{align}
where $\rho_{\rm g}$ is the local value of the galactic density \citep{HF00}. Exploiting the numerical values of the Oort's constants provided in \citet{FW97}, we obtain the dimensionless quantities $\kappa^2/\omega^2 \simeq 1.8$ and $\nu^2/\omega^2 \simeq 7.6$.

\section{Computational methods}
\label{cometh}

In order to study the escape mechanism of the orbits of the star cluster, we need to define samples of initial conditions of orbits whose properties (bounding or escaping motion) will be identified. The best method for this purpose, would have been to choose the sets of initial conditions of the orbits from a distribution function of the system. This however, is not available, so we define for each value of the energy (all tested energy levels are above the escape energy), dense uniform grids of initial conditions regularly distributed in the area allowed by the value of the energy. In three-dimensional (3D) systems, however, the phase space is six-dimensional and thus the behavior of the orbits cannot be easily visualized. One way to overcome this issue is to work in phase spaces with lower dimensions. Let us start with initial conditions on a 4D grid following the approach used successfully in \citet{ZCar13} and \citet{Z14}. In this way, we are able to identify regions of regularity/chaos and bound/escape, which may be visualized, if we restrict our investigation to a subspace of the whole 6D phase space. As in Paper I, our investigation takes place both in the physical $(x,y)$ and the phase $(x,\dot{x})$ space for a better understanding of the escape mechanism. In the physical $(x,y)$ space we consider orbits with initial conditions $(x_0, y_0, z_0)$, $\dot{y_0} = \dot{z_0} = 0$, while the initial value of $\dot{x_0}$ is obtained from the Jacobian integral (\ref{ham}). In particular, we define a value of $z_0$, which is kept constant and then we follow the evolution of the 3D orbits with initial conditions $(x_0, y_0)$, $\dot{y_0} = \dot{z_0} = 0$. Thus, we are able to construct a 2D plot depicting the physical $(x, y)$ plane but with an additional value of $z_0$, since we deal with 3D motion. All the initial conditions of the 3D orbits lie inside the limiting Zero Velocity Curve (ZVC) defined by
\begin{equation}
f_1(x,y;z_0) = \Phi_{\rm eff}(x, y, z=z_0) = E.
\label{zvc1}
\end{equation}
Similarly, for the phase $(x,\dot{x})$ space we consider orbits with initial conditions $(x_0, z_0, \dot{x_0})$, $y_0 = \dot{z_0} = 0$, while this time the initial value of $\dot{y_0}$ is obtained from the energy integral (\ref{ham}). Once more, the 3D orbits are launched with a specific value of $z_0$ and with initial conditions $(x_0, \dot{x_0})$, $y_0 = \dot{z_0} = 0$. The limiting curve which surrounds all the initial conditions of the 3D orbits and defines the allowed area on the 2D phase plane is
\begin{equation}
f_2(x,\dot{x};z_0) = \frac{1}{2}\dot{x}^2 + \Phi_{\rm eff}(x, y=0, z=z_0) = E.
\label{zvc2}
\end{equation}

In both cases, the step separation of the initial conditions along the $x$ and $y$ (physical space) and $x$ and $\dot{x}$ (phase space) axes (or in other words the density of the grids) was controlled in such a way that always there are about $10^5$ orbits, depending on the value of the Jacobian integral. The equations of motion as well as the variational equations for the initial conditions of all grids were integrated using a double precision Bulirsch-Stoer \verb!FORTRAN 77! algorithm \citep[e.g.,][]{PTVF92} with a small time step of order of $10^{-2}$, which is sufficient enough for the desired accuracy of our computations (i.e., our results practically do not change by halving the time step). Here we should emphasize, that our previous numerical experience suggests that the Bulirsch-Stoer integrator is both faster and more accurate than a double precision Runge-Kutta-Fehlberg algorithm of order 7 with Cash-Karp coefficients. Throughout all our computations, the Jacobian energy integral (Eq. (\ref{ham})) was conserved better than one part in $10^{-10}$, although for most orbits it was better than one part in $10^{-11}$.

In dynamical systems with escapes an issue of paramount importance is the determination of the position as well as the time at which an orbit escapes. For all values of energy smaller than the critical value $C_L$ (escape energy), the 3D equipotential surface is closed. On the other hand, when $E > C_L$ the equipotential surface is open and extend to infinity. The value of the energy itself however, does not furnish a sufficient condition for escape. An open equipotential surface consists of two branches forming channels through which an orbit can escape to infinity (see Fig. \ref{surf3d}). It was proved \citep{C79} that in Hamiltonian systems at every opening there is a highly unstable periodic orbit close to the line of maximum potential which is called a Lyapunov orbit. Such an orbit reaches the ZVC, on both sides of the opening and returns along the same path thus, connecting two opposite branches of the ZVC. Lyapunov orbits are very important for the escapes from the system, since if an orbit intersects any one of these orbits with velocity pointing outwards moves always outwards and eventually escapes from the system without any further intersections with the surface of section \citep[e.g.,][]{C90}. When $E = C_L$ the Lagrangian points exist precisely but at $E > C_L$ an unstable Lyapunov periodic orbit is located close to each of these two points \citep[e.g.,][]{H69}. The Lagrangian points are saddle points of the effective potential, so when $E > C_L$, a star must pass close enough to one of these points in order to escape. Thus, in the case of a tidally limited star cluster the escape criterion is purely geometric. In particular, escapers are defined to be those stars moving in 3D orbits beyond the tidal radius (1 pc) thus passing one of the two Lagrangian points \citep{FH00}.

\begin{figure}
\includegraphics[width=\hsize]{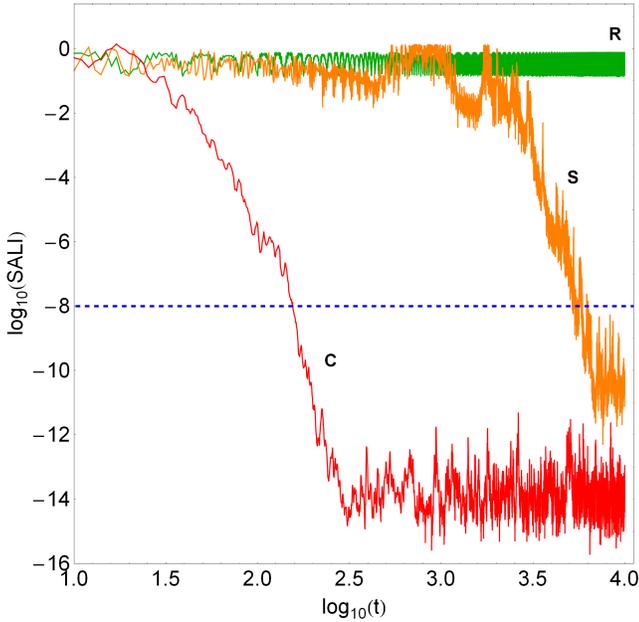}
\caption{Time-evolution of the SALI of a regular orbit (green color - R), a sticky orbit (orange color - S) and a chaotic orbit (red color - C) for a time period of $10^4$ time units. The horizontal, blue, dashed line corresponds to the threshold value $10^{-8}$ which separates regular from chaotic motion. The chaotic orbit needs only about 150 time units in order to cross the threshold, while on the other hand, the sticky orbit requires a considerable longer integration time of about 5000 time units so as to reveal its true chaotic nature.}
\label{salis}
\end{figure}

The physical and the phase space are divided into the escaping and non-escaping (trapped) space. Usually, the vast majority of the trapped space is occupied by initial conditions of regular orbits forming stability islands where a third, non-classical integral, also known as ``adelphic integral", is present. In many systems however, trapped chaotic orbits have also been observed. Therefore, we decided to distinguish between regular and chaotic trapped orbits. Over the years, several chaos indicators have been developed in order to determine the character of orbits. In our case, we chose to use the Smaller ALingment Index (SALI) method. The SALI \citep{S01} has been proved a very fast, reliable and effective tool, which is defined as
\begin{equation}
\rm SALI(t) \equiv min(d_-, d_+),
\label{sali}
\end{equation}
where $d_- \equiv \| {\bf{w_1}}(t) - {\bf{w_2}}(t) \|$ and $d_+ \equiv \| {\bf{w_1}}(t) + {\bf{w_2}}(t) \|$ are the alignments indices, while ${\bf{w_1}}(t)$ and ${\bf{w_2}}(t)$, are two deviation vectors which initially point in two random directions. For distinguishing between ordered and chaotic motion, all we have to do is to compute the SALI along time interval $t_{max}$ of numerical integration. In particular, we track simultaneously the time-evolution of the main orbit itself as well as the two deviation vectors ${\bf{w_1}}(t)$ and ${\bf{w_2}}(t)$ in order to compute the SALI.

The time-evolution of SALI strongly depends on the nature of the computed orbit since when an orbit is regular the SALI exhibits small fluctuations around non zero values, while on the other hand, in the case of chaotic orbits the SALI after a small transient period it tends exponentially to zero approaching the limit of the accuracy of the computer $(10^{-16})$. Therefore, the particular time-evolution of the SALI allow us to distinguish fast and safely between regular and chaotic motion. The time-evolution of a regular (R) and a chaotic (C) orbit for a time period of $10^4$ time units is presented in Fig. \ref{salis}. We observe, that both regular and chaotic orbits exhibit the expected behavior. Nevertheless, we have to define a specific numerical threshold value for determining the transition from order to chaos. After conducting extensive numerical experiments, integrating many sets of orbits, we conclude that a safe threshold value for the SALI is the value $10^{-8}$. The horizontal, blue, dashed line in Fig. \ref{salis} corresponds to that threshold value which separates regular from chaotic motion. In order to decide whether an orbit is regular or chaotic, one may follow the usual method according to which we check after a certain and predefined time interval of numerical integration, if the value of SALI has become less than the established threshold value. Therefore, if SALI $\leq 10^{-8}$ the orbit is chaotic, while if SALI $ > 10^{-8}$ the orbit is regular thus making the distinction between regular and chaotic motion clear and beyond any doubt. For the computation of SALI we used the \verb!LP-VI! code \citep{CMD14}, a fully operational routine which efficiently computes a suite of many chaos indicators for dynamical systems in any number of dimensions.

In our computations, we set $10^4$ time units (at the order of one Hubble time) as a maximum time of numerical integration. The vast majority of orbits (regular and chaotic) however, need considerable less time to find one of the two exits in the limiting surface and eventually escape from the system (obviously, the numerical integration is effectively ended when an orbit passes through one of the escape channels and escapes). Nevertheless, we decided to use such a vast integration time just to be sure that all orbits have enough time in order to escape. Remember, that there are the so called ``sticky orbits" which behave as regular ones during long periods of time. A characteristic example of a sticky orbit (S) can be seen in Fig. \ref{salis}, where we observe that the chaotic character of the particular sticky orbit is revealed only after a considerable long integration time of about 5000 time units. Here we should clarify, that orbits which do not escape after a numerical integration of $10^4$ time units (10 billion years) are considered as non-escaping or trapped. In fact, orbits with escape periods equal to many Hubble times are completely irrelevant to our investigation since they lack physical meaning.

\section{Numerical results}
\label{numres}

\begin{figure*}
\centering
\resizebox{\hsize}{!}{\includegraphics{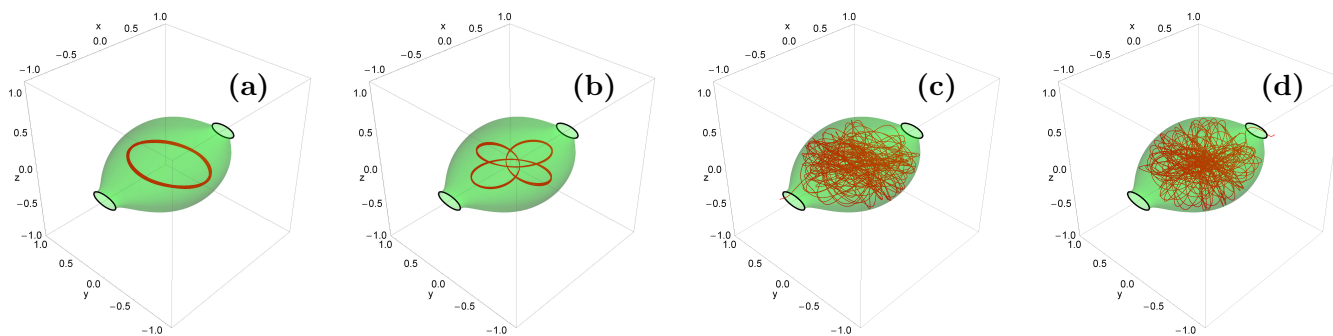}}
\caption{Characteristic examples of the main types of orbits in the star cluster. (a): 1:1:0 thin z-tube orbit, (b): 3:3:0 resonant orbit, (c): orbit escaping through $L_1$, (d): orbit escaping through $L_2$.}
\label{orbs}
\end{figure*}

The main objective of our investigation is to determine which orbits escape and which remain trapped, distinguishing simultaneously between regular and chaotic trapped motion\footnote{Generally, any dynamical method requires a sufficient time interval of numerical integration in order to distinguish safely between ordered and chaotic motion. Therefore, if the escape rate of an orbit is very low or even worse if the orbit escapes directly from the system then, any chaos indicator (the SALI in our case) will fail to work properly due to insufficient integration time. Hence, it is pointless to speak of regular or chaotic escaping orbits.}. Furthermore, two additional properties of the orbits will be examined: (i) the channels through which the stars escape and (ii) the time-scale of the escapes (we shall also use the terms escape period or escape rates). In the present paper, we shall explore these dynamical quantities for various values of the energy, always within the interval $\widehat{C} \in [0.001,0.1]$, as well as for the initial value of the $z$ coordinate. In particular, three different cases are considered: (a) orbits starting very close to the $(x,y)$ plane with low value of $z_0$, (b) orbits having a intermediate initial value of $z_0$ and (c) orbits launched at relatively large distance from the $(x,y)$ plane thus with high value of $z_0$.

Our numerical calculations indicate that apart from the escaping orbits there is always a considerable amount of non-escaping orbits. In general terms, the majority of non-escaping regions corresponds to initial conditions of regular orbits, where the adelphic integral of motion is present, restricting their accessible phase space and therefore hinders their escape. However, there are also chaotic orbits which do not escape within the predefined interval of $10^4$ time units and remain trapped for vast periods (even up to 100 Hubble times) until they eventually escape to infinity. Therefore, we decided to classify the initial conditions of orbits in both the physical and phase space into four main categories: (i) orbits that escape through $L_1$, (ii) orbits that escape through $L_2$, (iii) non-escaping regular orbits and (iv) trapped chaotic orbits. Additional numerical computations reveal that the non-escaping regular orbits are mainly 3D tube orbits for which the adelphic integral applies\footnote{The total angular momentum is an approximately conserved quantity (integral of motion), even for orbits in non spherical potentials.}, while other types of secondary resonant orbits are also present. In Fig. \ref{orbs}a we present an example of a 1:1:0 thin $z$-tube orbit, while in Fig. \ref{orbs}b a characteristic example of a secondary 3:3:0 resonant orbit is shown. The $n:m:l$ notation we use for the regular orbits is according to \citet{CA98} and \citet{ZC13}, where the ratio of those integers corresponds to the ratio of the main frequencies of the orbit, where main frequency is the frequency of greatest amplitude in each coordinate. Main amplitudes, when having a rational ratio, define the resonances of an orbit. Finally in Figs. \ref{orbs}c-d we observe two orbits escaping through $L_1$ (exit channel 1) and $L_2$ (exit channel 2), respectively. Both regular orbits shown in Figs. \ref{orbs}a-b are computed until $t = 100$ time units, while on the other hand the escaping orbits presented in Figs. \ref{orbs}c-d were calculated for 5 time units more than the corresponding escape period in order to visualize better the escape trail.

\subsection{Orbits starting with a low value of $z_0$}
\label{case1}

\begin{figure*}
\centering
\resizebox{0.8\hsize}{!}{\includegraphics{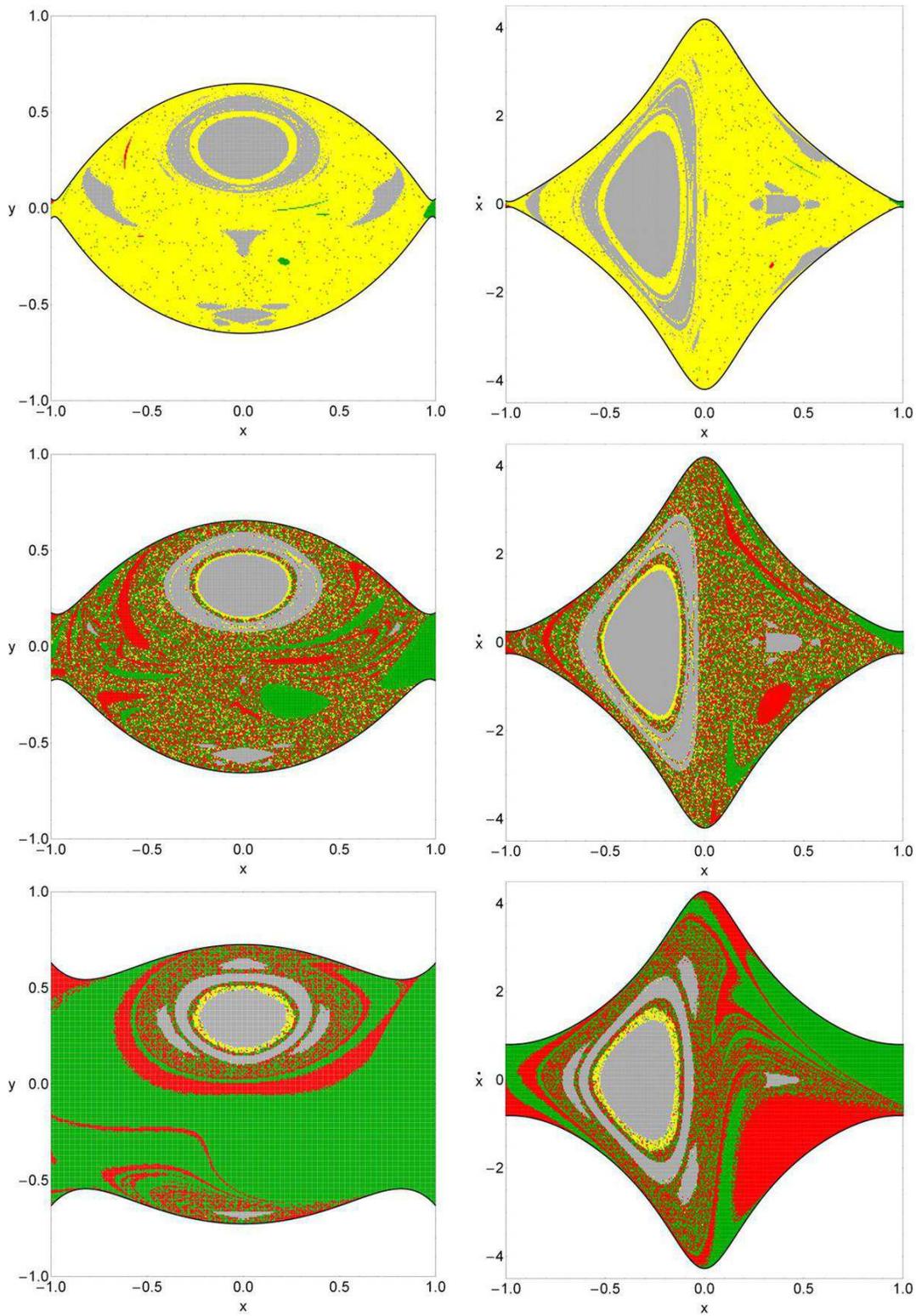}}
\caption{Orbital structure of the physical $(x,y)$ and phase $(x,\dot{x})$ space of 3D orbits with $z_0 = 0.01$. Top row: $\widehat{C} = 0.001$; Middle row: $\widehat{C} = 0.01$; Bottom row: $\widehat{C} = 0.1$. The red regions correspond to initial conditions of orbits where the stars escape through $L_1$, green regions denote initial conditions where the stars escape through $L_2$, gray areas represent stability islands of regular non-escaping orbits, while initial conditions of trapped chaotic orbits are marked in yellow.}
\label{grd1}
\end{figure*}

\begin{figure*}
\centering
\resizebox{0.8\hsize}{!}{\includegraphics{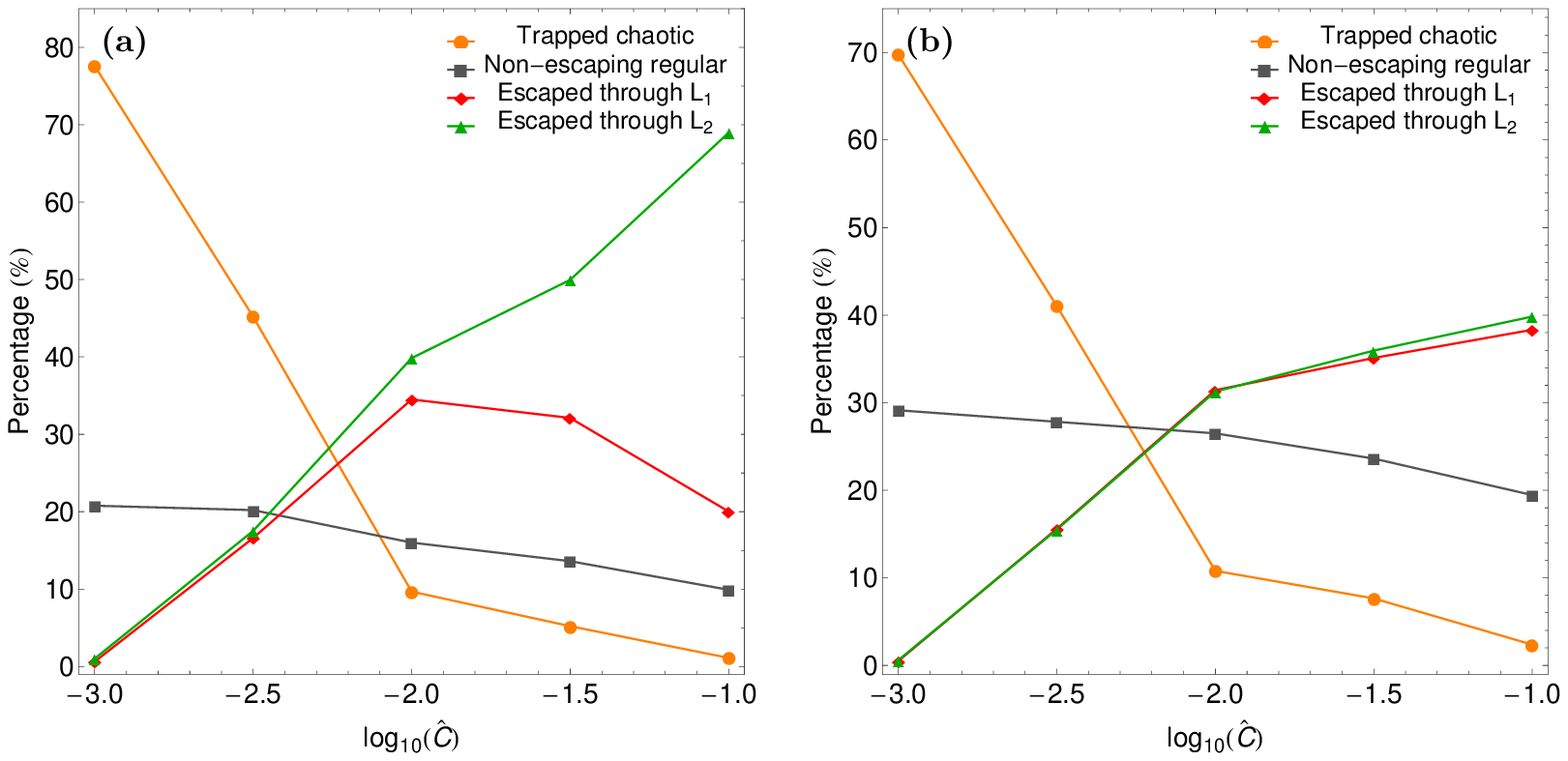}}
\caption{Evolution of the percentages of trapped, escaping and non-escaping orbits when varying the energy parameter $\widehat{C}$ for $z_0 = 0.01$ on the (a-left): physical $(x,y)$ space and (b-right): phase $(x,\dot{x})$ space.}
\label{percs1}
\end{figure*}

\begin{figure*}
\centering
\resizebox{0.8\hsize}{!}{\includegraphics{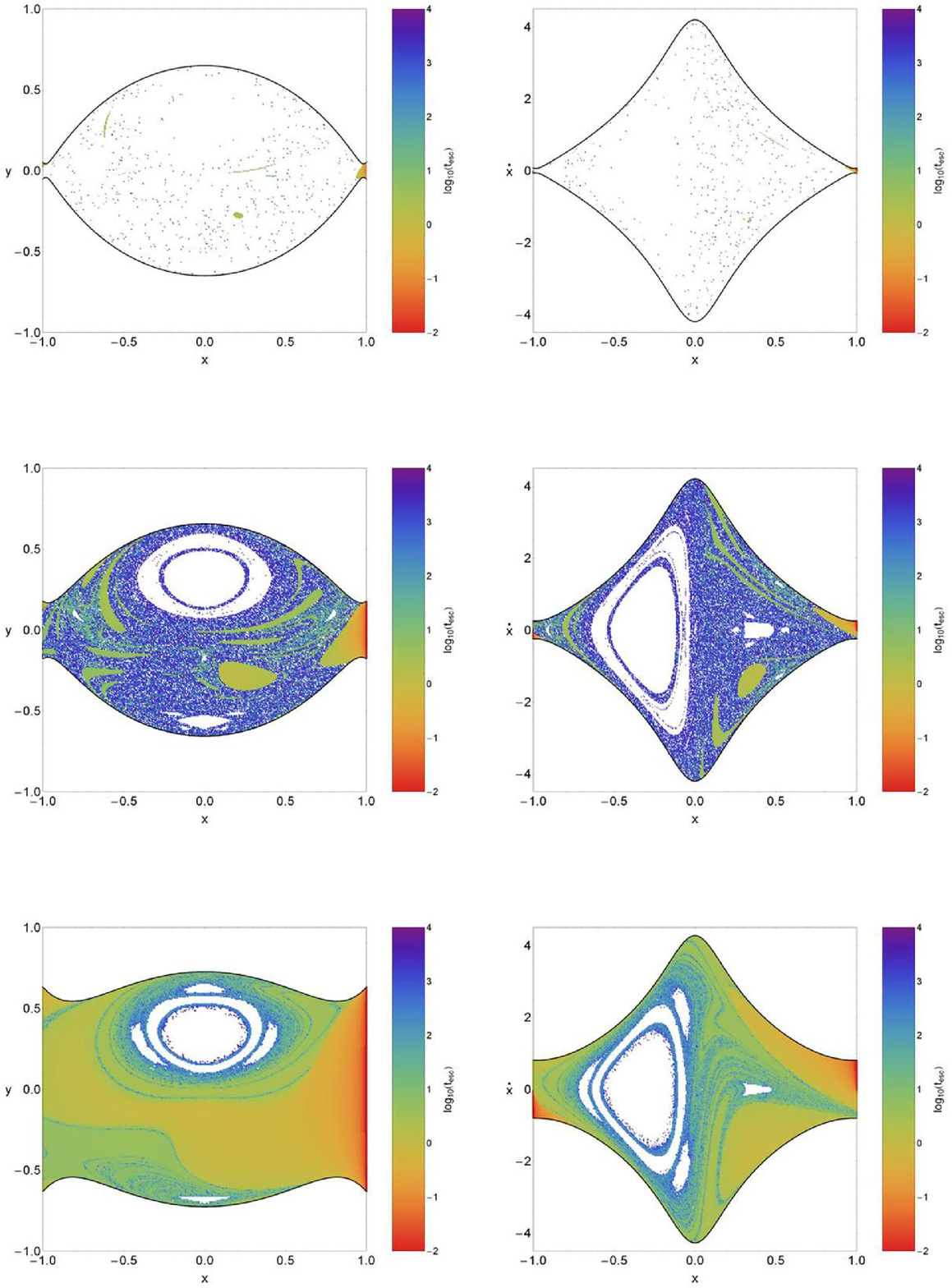}}
\caption{Distribution of the escape times $t_{\rm esc}$ of the orbits on the physical $(x,y)$ and phase $(x,\dot{x})$ space when $z_0 = 0.01$. Top row: $\widehat{C} = 0.001$; Middle row: $\widehat{C} = 0.01$; Bottom row: $\widehat{C} = 0.1$. The darker the color, the larger the escape time. Initial conditions of non-escaping regular orbits and trapped chaotic orbits are shown in white.}
\label{tesc1}
\end{figure*}

Our exploration begins considering 3D orbits starting very close to the primary $(x,y)$ plane with initial condition $z_0 = 0.01$. Fig. \ref{grd1} shows the structure of the physical $(x,y)$ and phase $(x,\dot{x})$ space for several values of the energy, where the four basic types of orbits are indicated with different colors. Specifically, gray color corresponds to regular non-escaping orbits, yellow color corresponds to trapped chaotic orbits, red color corresponds to orbits escaping through channel 1, while the initial conditions of orbits escaping through exit channel 2 are marked with green color. The outermost black solid lines in the two types of plots denote the limiting curves defined by Eq. (\ref{zvc1}) and Eq. (\ref{zvc2}), respectively. Here we should note that since $\Phi_{\rm eff}$ is mirror-symmetrical to all axes, the asymmetry (also known as ``Coriolis asymmetry") seen in the plots is due to the presence of the Coriolis forces \citep{I80}. We observe, that our results are very similar with those of the 2D system presented in Fig. 5 of Paper I however, there are also major differences. In particular, it can be seen that for all three energy levels the stability regions with initial conditions of non-escaping regular orbits are almost identical with those of the 2D system\footnote{Remember that in Paper I there was no classification between ordered and chaotic motion thus both non-escaping regular orbits and trapped chaotic ones were put together using the same identification color (black).}. The regular areas correspond mainly to retrograde orbits \citep[see also][]{FH00,EGFJ07} (i.e., when a star revolves around the cluster in the opposite sense with respect to the motion of the cluster around the parent galaxy), while there are also some smaller stability islands of prograde orbits. Similarities for the basins of escape when $\widehat{C} = 0.01$ and $\widehat{C} = 0.1$ with the corresponding results of Paper I are also observed. The most noticeable difference on the structure of both types of planes on the other hand, is the amount of trapped chaotic orbits. For $\widehat{C} = 0.001$, that is an energy level just above the critical threshold value $C_L$, we see that the vast majority of initial conditions correspond to trapped chaotic orbits, while only a tiny portion manage to escape within the predefined time interval of numerical integration. As we proceed to higher energy levels however, the amount of trapped chaotic orbits reduces significantly. At the same time, the phase space is occupied by several exit regions where we observe a high dependence of the escape mechanism on the particular initial conditions of the orbits. In other words, a minor change in the initial conditions has as a result the star to escape through the opposite escape channel, which is of course a classical indication of chaotic motion. Moreover, with increasing energy the area in the phase spaces becomes less and less fractal and well-defined basins of escape emerge, where by the term basin of escape, we refer to a local set of initial conditions that corresponds to a certain escape channel. We would like to point that for $\widehat{C} = 0.01$ and $\widehat{C} = 0.1$, in both the physical and phase space, we identify the existence of thin rings of escaping orbits inside the main stability island of regular orbits. This behaviour however, was not observed in the 2D model of Paper I. We are convinced that these rings composed of initial conditions of escaping are connected to the outer world at high enough values of $z$ (see also Figs. \ref{xc} and \ref{xz}).

In Fig. \ref{percs1}(a-b) we present the evolution of the percentages of all types of orbits on the physical and phase space when the dimensionless energy parameter $\widehat{C}$ varies. One may observe, that when $\widehat{C} = 0.001$, that is just above the critical energy level $C_L$, escaping orbits are practical absent (with rates less than 1\%), while both the physical and phase spaces are covered almost entirely with initial conditions of bounded orbits (non-escaping regular plus trapped chaotic). In particular, about 80\% of the physical space and 20\% of the phase space corresponds to trapped chaotic orbits, while the remaining is occupied by stability islands of non-escaping regular orbits. As the value of energy increases however, the percentage of trapped motion reduces, while at the same time, the amount of escaping orbits grows steadily. The rate of trapped chaotic orbits drops rapidly in the interval $\widehat{C} \in [0.001,0.01]$ and at the highest energy level studied $(\widehat{C} = 0.1)$ it vanishes. On the other hand, the percentage of regular non-escaping orbits is much less affected by the shifting of the energy displaying a slow but constant reduction roughly from 20\% to 10\% and from 30\% to 20\% in the physical and phase space, respectively. The evolution of the rates of escaping orbits is the same in both planes only for $\widehat{C} \leq 0.01$, where the amount of escaping orbits through exits 1 and 2 increases linearly and almost identically. In the physical space for $\widehat{C} > 0.01$ we see that the percentage of escaping orbits through $L_1$ exhibits a decrease and for $\widehat{C} = 0.1$ they occupy only about one fifth of the plane, while the rate of escaping orbits through $L_2$ continuous to grow and for $\widehat{C} = 0.1$ they dominate covering about 70\% of the physical space. On the contrary, in the phase space it is seen that the amount of escaping orbits is still increasing linearly, although with smaller slope than before, and the percentages of both exit channels are almost the same reaching about 40\% at the highest energy level studied.

The following Fig. \ref{tesc1} shows how the escape times $t_{\rm esc}$ of orbits are distributed on both the physical $(x,y)$ and phase $(x,\dot{x})$ space. Light reddish colors correspond to fast escaping orbits with short escape periods, dark blue/purple colors indicate large escape rates, while white color denote both non-escaping regular and trapped chaotic orbits. It is evident, that orbits with initial conditions close to the boundaries between the escape basins, that is, the fractal area of the plots need significant amount of time in order to escape from the cluster, while on the other hand, inside the basins of escape where there is no dependence on the initial conditions whatsoever, we measured the shortest escape rates of the orbits. We observe that for $\widehat{C} = 0.001$ almost all the available area inside the limiting curves is white since as we seen in Fig. \ref{grd1} the vast majority of orbits are either non-escaping regular or trapped chaotic. The pattern however, changes drastically for $\widehat{C} = 0.01$ where it can be seen that the escape periods of orbits with initial conditions in the fractal basin boundaries are huge corresponding to tens of thousands of time units. This phenomenon is anticipated because in this case the width of the escape channels is still relatively small and therefore, the orbits should spend much time inside the equipotential surface until they find one of the two openings and eventually escape to infinity. One may observe, that as the value of the energy increases however, the escape channels become more and more wide, while the escape times of all orbits (inside the exit basins and in the fractal regions) is reduced considerably. In general terms, the behavior of the escape times of star orbits is quite similar to that reported in Paper I, where only the 2D case was studied.

\subsection{Orbits starting with an intermediate value of $z_0$}
\label{case2}

\begin{figure*}
\centering
\resizebox{0.8\hsize}{!}{\includegraphics{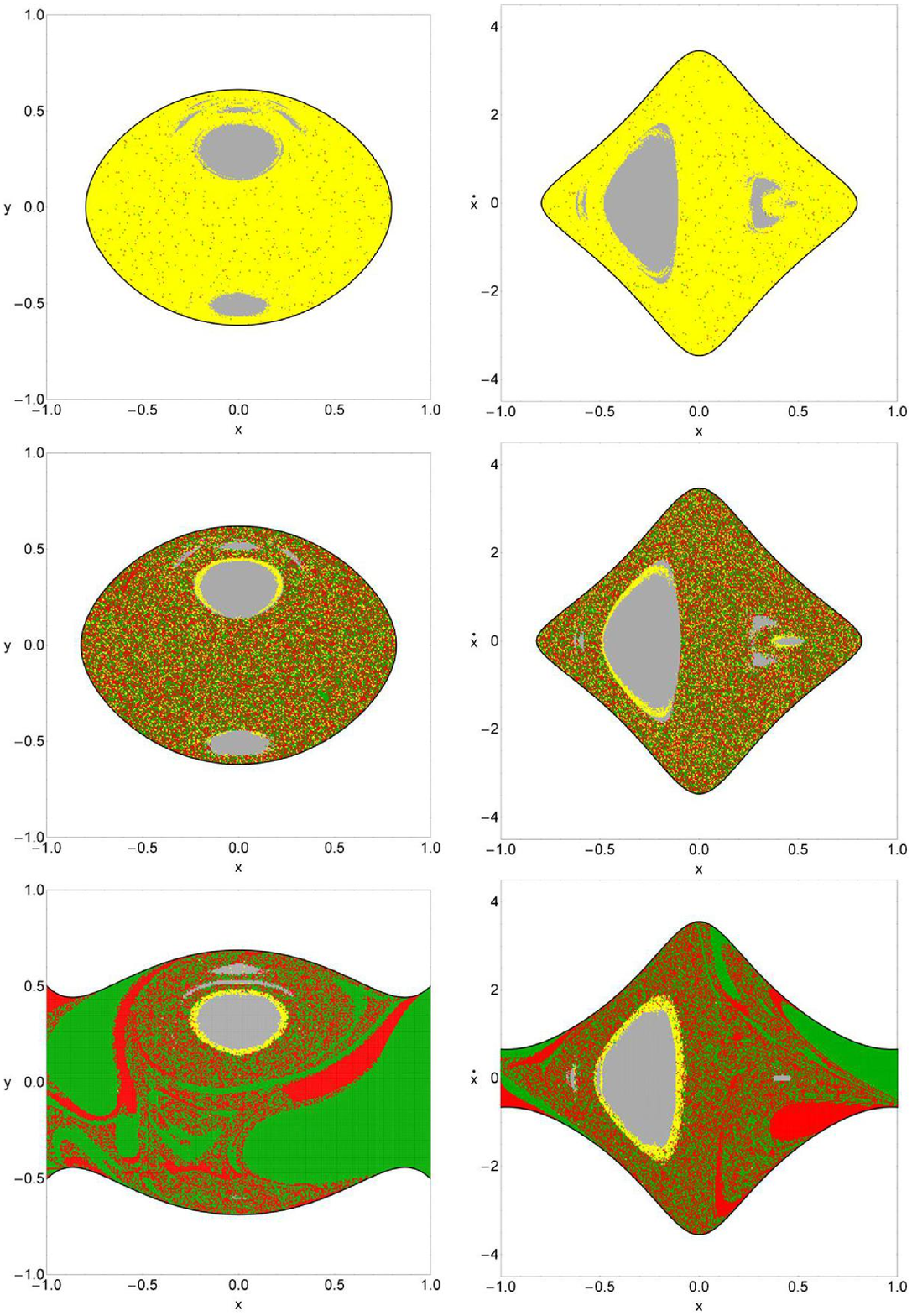}}
\caption{Orbital structure of the physical $(x,y)$ and phase $(x,\dot{x})$ space of 3D orbits with $z_0 = 0.15$. Top row: $\widehat{C} = 0.001$; Middle row: $\widehat{C} = 0.01$; Bottom row: $\widehat{C} = 0.1$. The color code is the same as in Fig. \ref{grd1}.}
\label{grd2}
\end{figure*}

\begin{figure*}
\centering
\resizebox{0.8\hsize}{!}{\includegraphics{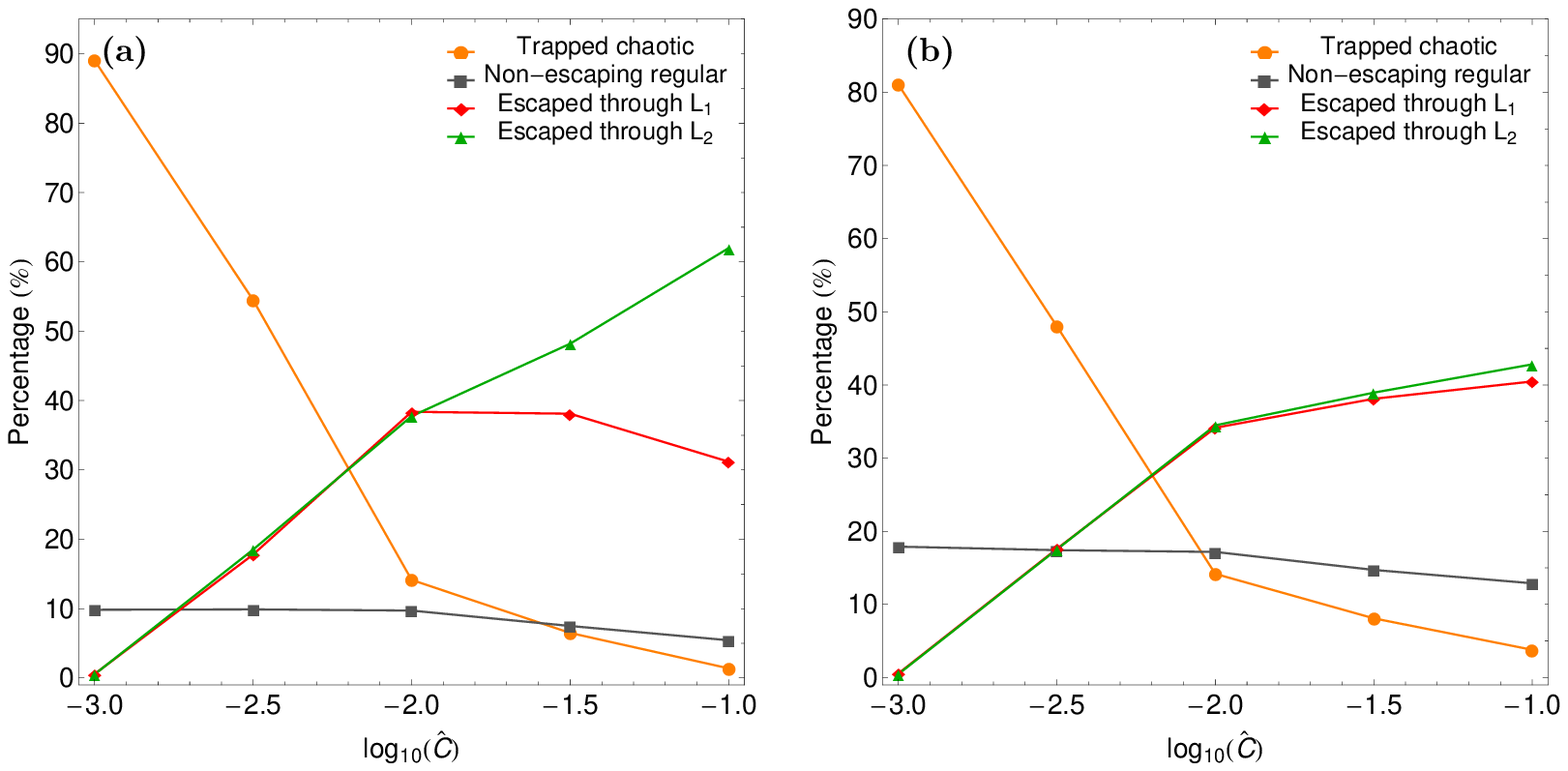}}
\caption{Evolution of the percentages of trapped, escaping and non-escaping orbits when varying the energy parameter $\widehat{C}$ for $z_0 = 0.15$ on the (a-left): physical $(x,y)$ space and (b-right): phase $(x,\dot{x})$ space.}
\label{percs2}
\end{figure*}

\begin{figure*}
\centering
\resizebox{0.8\hsize}{!}{\includegraphics{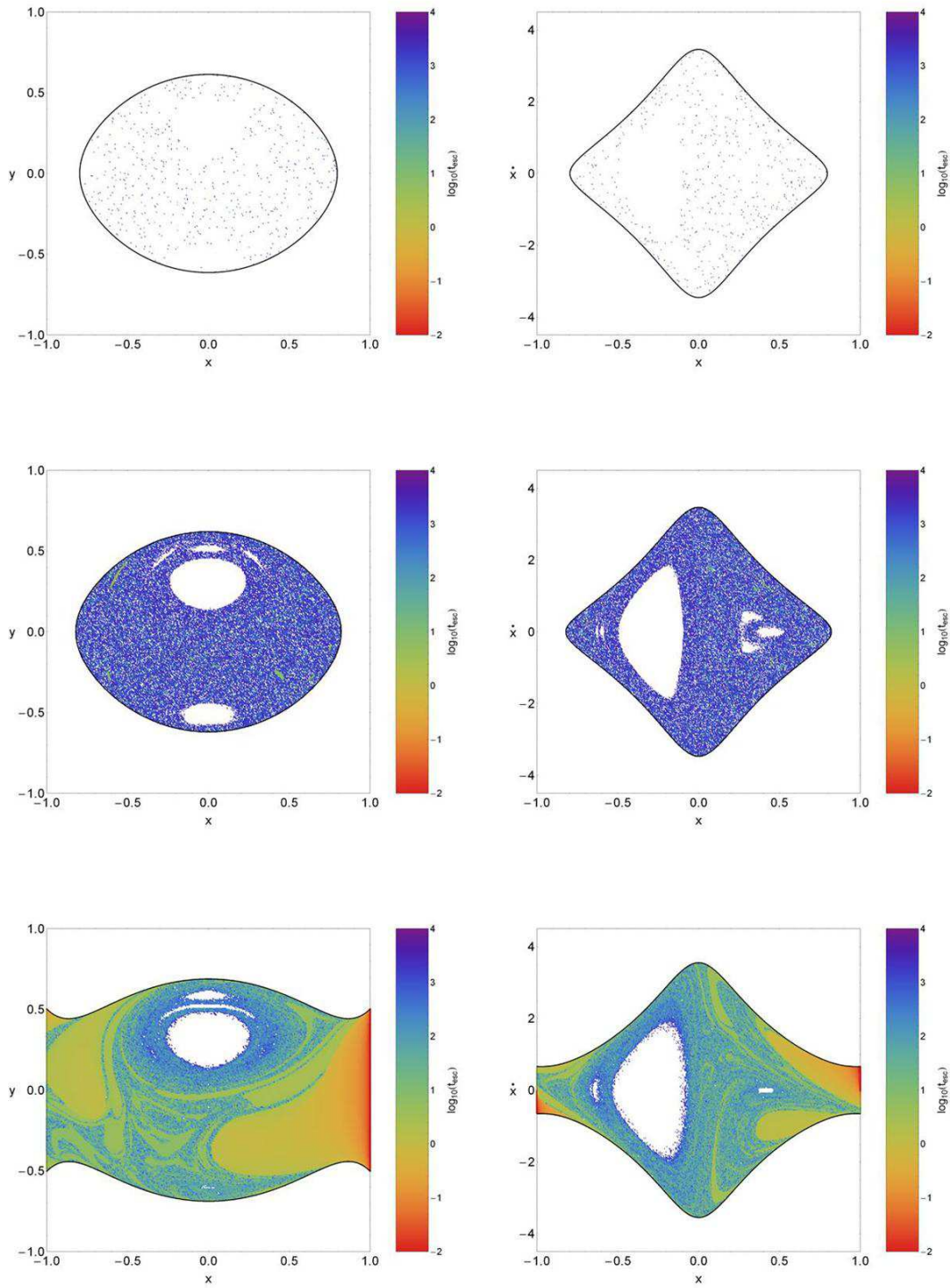}}
\caption{Distribution of the escape times $t_{\rm esc}$ of the orbits on the physical $(x,y)$ and phase $(x,\dot{x})$ space when $z_0 = 0.15$. Top row: $\widehat{C} = 0.001$; Middle row: $\widehat{C} = 0.01$; Bottom row: $\widehat{C} = 0.1$. The color code is the same as in Fig. \ref{tesc1}.}
\label{tesc2}
\end{figure*}

We continue our quest considering in this case three-dimensional orbits starting at a moderate distance from the primary $(x,y)$ plane having initial condition $z_0 = 0.15$ and we follow the same numerical approach as previously. In Fig. \ref{grd2} we present the structure of the physical and phase planes for three energy levels, using different colors in order to distinguish between the four main types of orbits (non-escaping regular; trapped chaotic; escaping through $L_1$ and escaping through $L_2$). The results are quite similar to that discussed previously in Fig. \ref{grd1} nevertheless, there are a few noticeable differences. In particular we observe that the increase of the initial value of the $z$ coordinate of orbits, has affected the geometry of both types of planes, since the available space has been reduced. Furthermore, for $\widehat{C} = 0.001$ and $\widehat{C} = 0.01$ we see that the limiting curves are closed, this however, does not imply by no means that escape is impossible. To understand why the limiting curves are closed, we have to look again carefully the shape of the 3D equipotential surface shown in Fig. \ref{surf3d} and observe that the two openings (exit throats) are at relatively low distance above the primary $(x,y)$ plane. The $z_0 = 0.15$ slice is simply above the exit channels and therefore, the limiting curves appear to be closed showing no signs of exits. We will further discuss and also explain this issue later in the next section. Once more it is seen that when $\widehat{C} = 0.001$ trapped chaotic motion is the dominant type of motion among the orbits. This behaviour changes drastically for $\widehat{C} = 0.01$, where trapped chaotic orbits are confined mainly to the boundaries of the stability islands, while the remaining area is completely fractal (escape basins, if any, are negligible) and is occupied by initial conditions of escaping orbits. The degree of fractalization however, reduces significantly for $\widehat{C} = 0.1$, where several well-defined basins of escape are present and the two openings appear themselves again. Here we have to point out that the stability islands corresponding to higher resonant orbits look less prominent with respect to that we seen in Fig. \ref{grd1}, where $z_0 = 0.01$. The stability islands of the main 3D tube orbits on the other hand, seem to be retained. The distribution of the escape times $t_{\rm esc}$ of orbits on both the physical and phase space is shown in Fig. \ref{tesc2}. One may observe that the results are very similar to those presented earlier in Fig. \ref{tesc1}, where we found that orbits with initial conditions inside the escape basins have the smallest escape rates, while on the other hand, the longest escape times correspond to orbits with initial conditions in the fractal regions of the plots.

The evolution of the percentages of all types of orbits on the physical and phase space as a function of the dimensionless energy parameter $\widehat{C}$ is presented in Fig. \ref{percs2}(a-b). We see that once more, for $\widehat{C} = 0.001$ the vast majority of the initial conditions (about 90\% of the physical and 80\% of the phase scape) corresponds either to trapped chaotic or non-escaping regular orbits, while there is almost zero amount of escaping orbits. The rate of trapped chaotic orbits drops drastically with increasing energy and eventually vanishes at $\widehat{C} = 0.1$, while at the same time a gradual increase at the percentage of escaping orbits is observed. The percentage of ordered non-escaping orbits on the other hand, seems to be unaffected by the energy shifting, holding throughout about 8\% and 17\% of the physical and phase space, respectively. The rate of escaping orbits increases linearly in the interval $\widehat{C} \in [0.001,0.01]$, while for $\widehat{C} \geq 0.01$ escaping orbits is the most populated family in both planes. It is seen that in the physical space for $\widehat{C} > 0.01$ the percentages of escaping orbits through exit channels 1 and 2 start to diverge. In particular, the rate of escaping orbits through $L_1$ display a minor decrease, covering about 30\% of the physical space for $\widehat{C} = 0.1$, while escaping orbits through $L_2$ continue to increase their amount and at the highest energy level $(\widehat{C} = 0.1)$ they occupy twice the region of escaping orbits trough exit channel 1. In the phase space however, the percentages of escaping orbits do not diverge, they retain their mutual linear increase and for $\widehat{C} = 0.1$ they share about 80\% of the phase space. At this point it should be emphasized that the fact that the amount of escaping orbits of both channels is almost the same throughout the energy range implies that the two exit channels in the phase space can be considered equiprobable. This is also true in the physical space but only for $\widehat{C} \leq 0.01$.

\subsection{Orbits starting with a high value of $z_0$}
\label{case3}

\begin{figure*}
\centering
\resizebox{0.8\hsize}{!}{\includegraphics{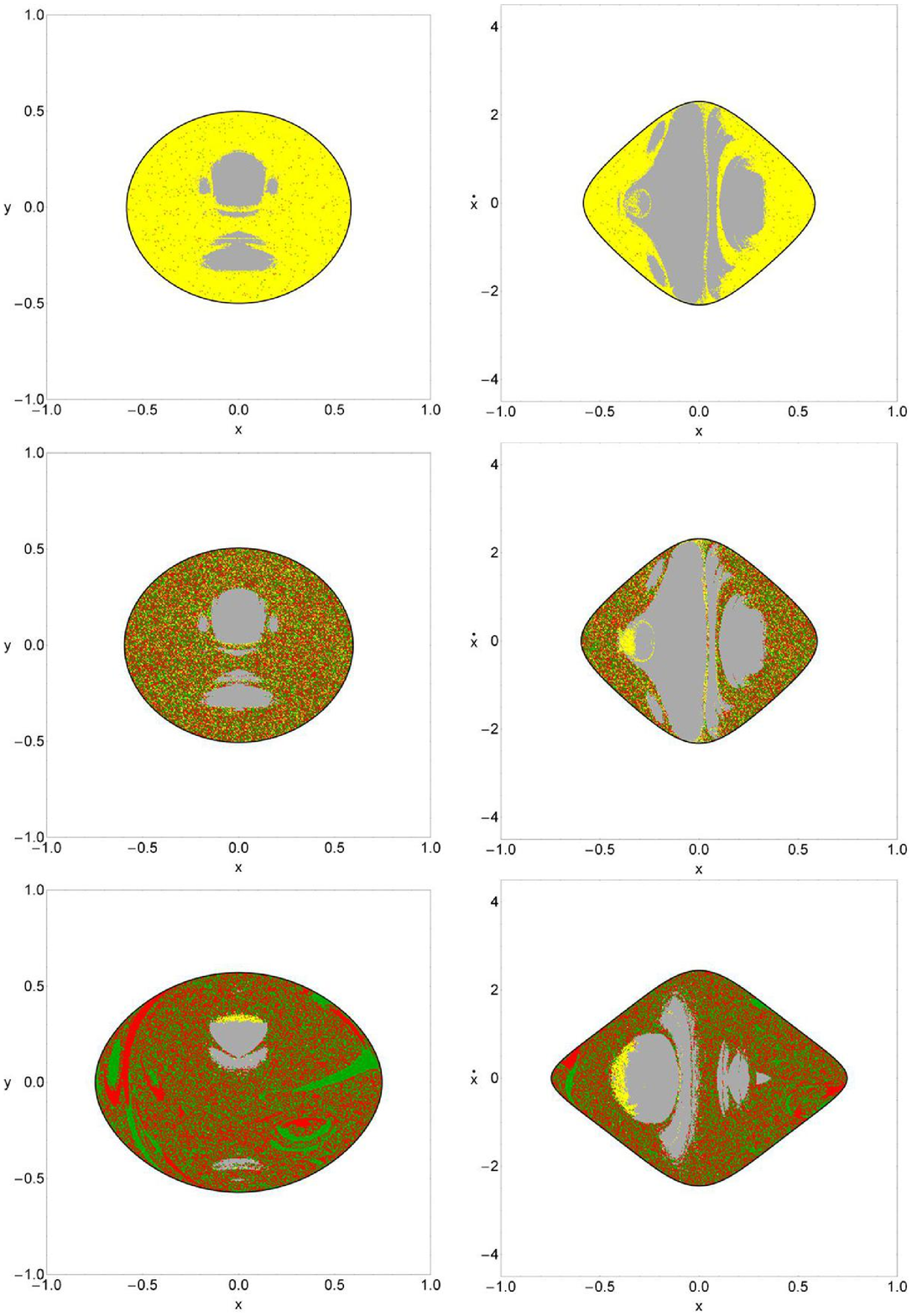}}
\caption{Orbital structure of the physical $(x,y)$ and phase $(x,\dot{x})$ space of 3D orbits with $z_0 = 0.3$. Top row: $\widehat{C} = 0.001$; Middle row: $\widehat{C} = 0.01$; Bottom row: $\widehat{C} = 0.1$. The color code is the same as in Fig. \ref{grd1}.}
\label{grd3}
\end{figure*}

\begin{figure*}
\centering
\resizebox{0.8\hsize}{!}{\includegraphics{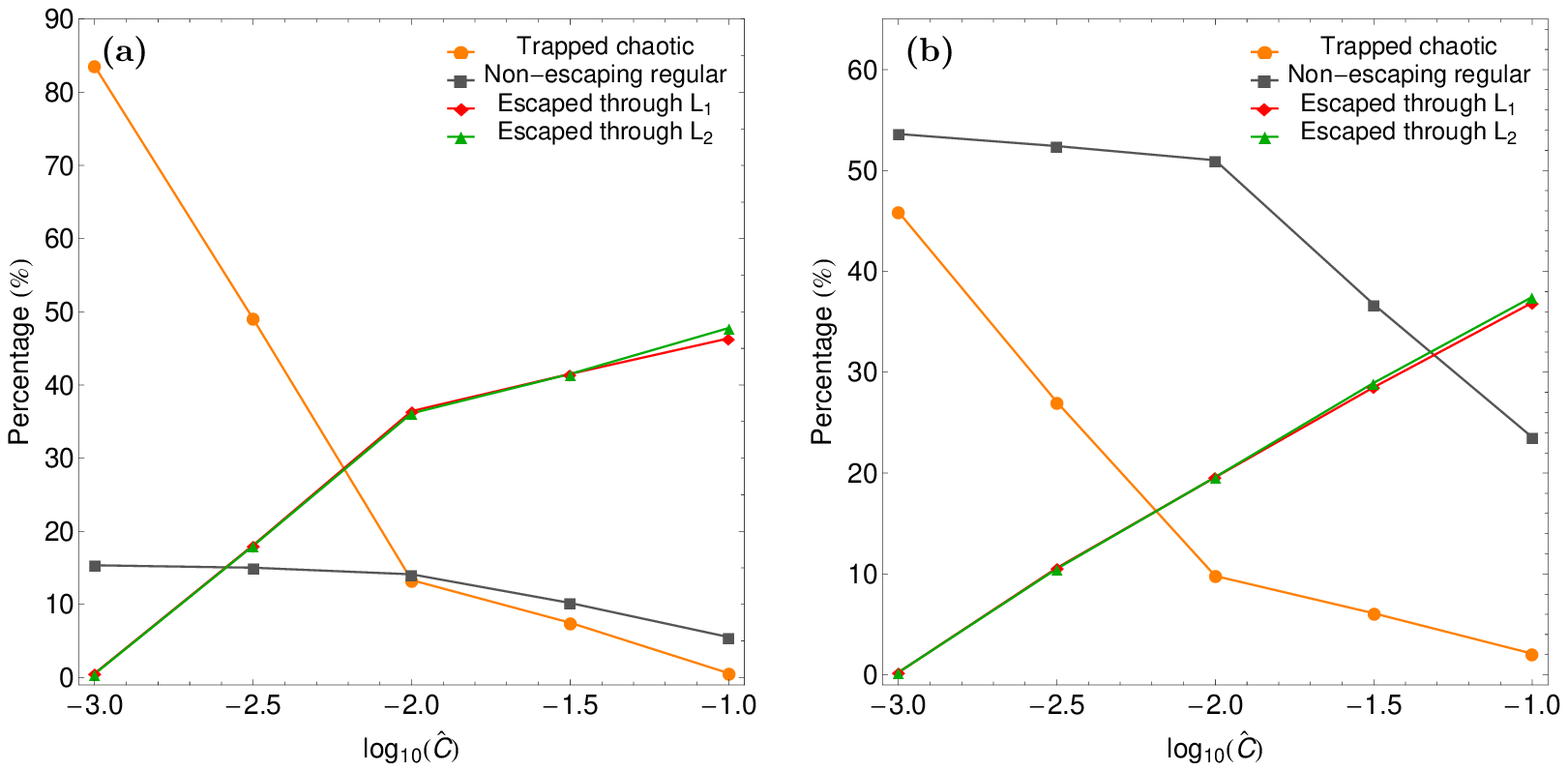}}
\caption{Evolution of the percentages of trapped, escaping and non-escaping orbits when varying the energy parameter $\widehat{C}$ for $z_0 = 0.3$ on the (a-left): physical $(x,y)$ space and (b-right): phase $(x,\dot{x})$ space.}
\label{percs3}
\end{figure*}

\begin{figure*}
\centering
\resizebox{0.8\hsize}{!}{\includegraphics{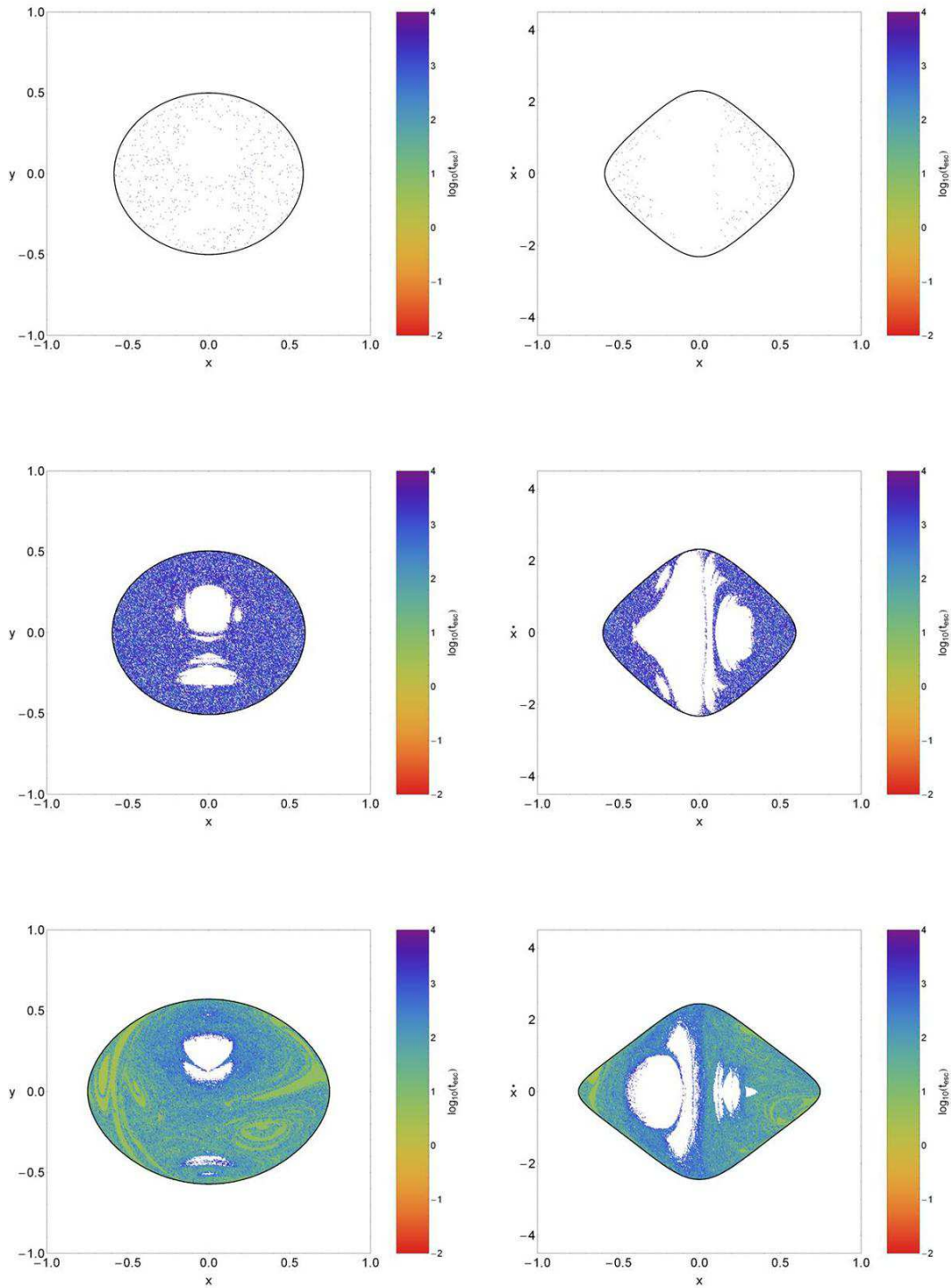}}
\caption{Distribution of the escape times $t_{\rm esc}$ of the orbits on the physical $(x,y)$ and phase $(x,\dot{x})$ space when $z_0 = 0.3$. Top row: $\widehat{C} = 0.001$; Middle row: $\widehat{C} = 0.01$; Bottom row: $\widehat{C} = 0.1$. The color code is the same as in Fig. \ref{tesc1}.}
\label{tesc3}
\end{figure*}

The last case under investigation involves three-dimensional orbits launched at a relatively high distance from the primary $(x,y)$ plane with initial condition $z_0 = 0.3$. Again, all the different aspects of the numerical approach remain exactly the same as in the two previously studied cases. The orbital structure of the physical and phase planes distinguishing between the four main types of orbits for three energy levels is shown in Fig. \ref{grd3}. It is seen that the allowed area of motion has been reduced even further with respect to that of Figs. \ref{grd1} and \ref{grd2}, while both limiting curves are closed in all three energy levels. Furthermore, we observe that the shape of the stability regions differs a lot from that of the previous cases, while sets of small stability islands corresponding to secondary resonances are still present. In particular, at $\widehat{C} = 0.001$ we identify several stability islands which are located mainly in the central region of the physical plane. In the phase space on the other hand, the regions of non-escaping ordered orbits are more extended and they occupy about half of the phase plane. The remaining space is filled by initial conditions of trapped chaotic orbits, while the few initial conditions of escaping orbits are randomly scattered all over the vast sea of trapped chaotic motion. The pattern regarding the stability islands remains almost the same at $\widehat{C} = 0.01$ but the structure of the rest space changes significantly. Specifically, the amount of trapped chaotic orbits is heavily reduced (they are mainly confined near the boundaries of stability islands) and both types of planes display a high degree of fractalization regarding the escaping orbits, without any indication of basins of escape. At $\widehat{C} = 0.1$ on the other hand, the area on the planes corresponding to non-escaping regular orbits reduces considerably and several thin, elongated basins of escape emerge mainly at the outer parts of both planes which however, still remain highly fractal. In Fig. \ref{tesc3} we depict the distribution of the escape times $t_{\rm esc}$ of orbits on both the physical and phase space, where one can see similar outcomes with that presented in the previous subsections. At this point, we would like to point out that the basins of escape can be easily distinguished in the two grids of the last row of Fig. \ref{tesc3}, being the regions with intermediate greenish colors indicating fast escaping orbits. Indeed, our numerical calculations suggest that orbits with initial conditions inside the basins escape from the star cluster after about 10 time units.

Finally, Fig. \ref{percs3}(a-b) shows how the percentages of all types of orbits on the physical and phase space evolve when the energy parameter varies in the interval $\widehat{C} \in [0.001,0.1]$. It is seen that for $\widehat{C} = 0.001$ about 85\% of the physical space is covered by initial conditions corresponding to trapped chaotic orbits, while only 45\% of the phase space is occupied by the same type of orbits. In particular, non-escaping regular orbits is the dominant type of orbits covering more than half of the phase space for $\widehat{C} \leq 0.01$, while for larger values of energy the percentage drops up to about 25\% for $\widehat{C} = 0.1$. In the physical space on the other hand, trapped chaotic orbits is the most populated family for small energy levels, while the rate of non-escaping orbits is much less influenced by the energy shifting reducing from about 15\% for $\widehat{C} = 0.001$ to about 5\% for $\widehat{C} = 0.1$. We also observe that just above the escape energy the amount of escaping orbits is practically zero (less than 1\%), with increasing energy however, their percentage displays a constant linear increase in both planes. In the physical $(x,y)$ plane the increase becomes less rapid for $\widehat{C} > 0.01$, as the slope decreases, and at the highest energy level studied escaping orbits take over occupying more than 90\% of the available physical space. On the contrary, in the phase $(x,\dot{x})$ plane the rapid linear increase of the percentages of escaping orbits remains constant throughout the entire range of the energy levels and for $\widehat{C} = 0.1$ they share about 74\% of the phase space. Looking the diagrams shown in Fig. \ref{percs3}(a-b) we may conclude that when orbits are started with a high initial value of $z_0$, the two escape channels can be considered equiprobable in both types of planes, taking into account that the rates of escaping orbits of both channels is almost the same throughout.

\subsection{An overview analysis}
\label{geno}

\begin{figure*}
\centering
\resizebox{0.8\hsize}{!}{\includegraphics{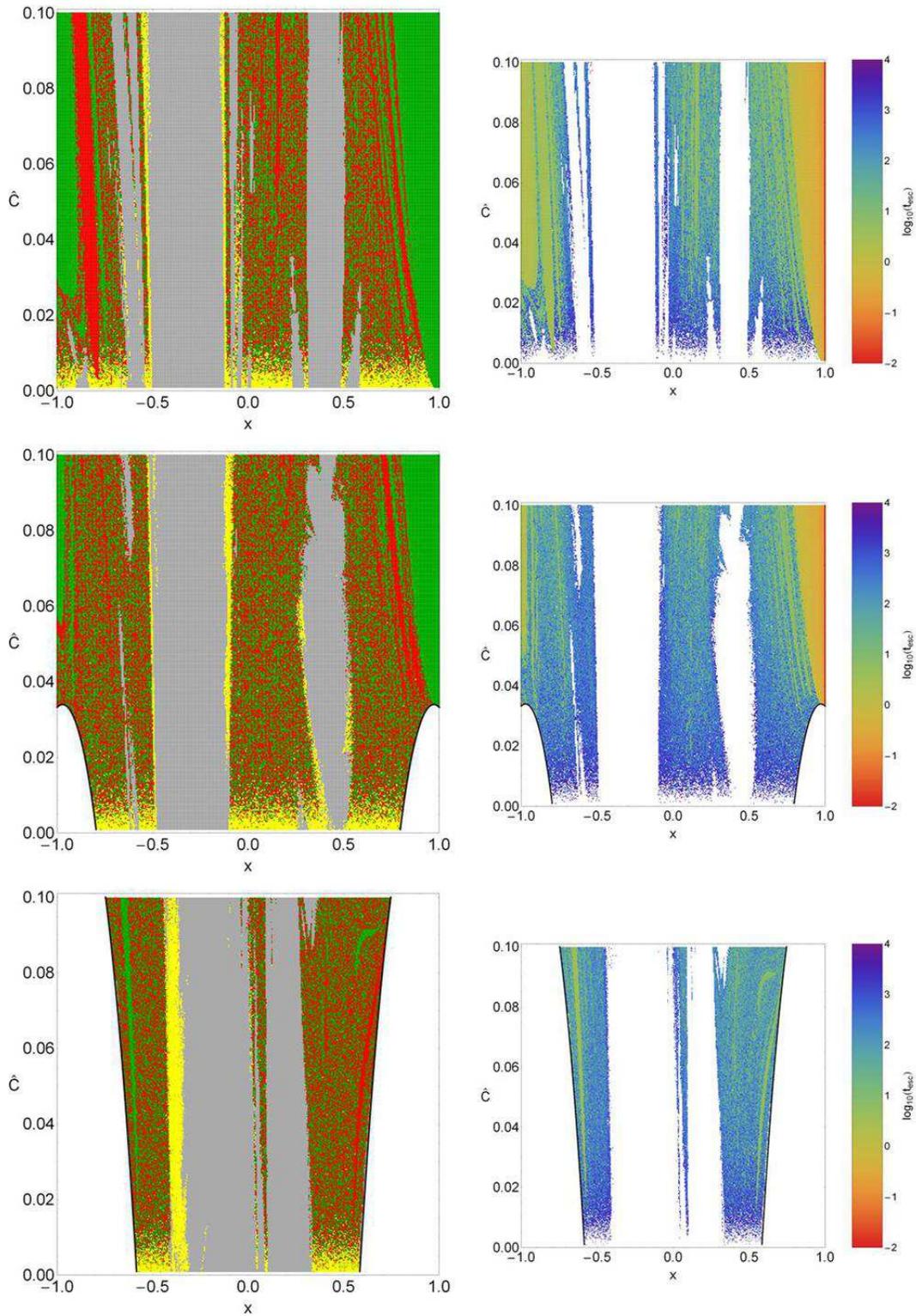}}
\caption{(left column): Orbital structure of the $(x,\widehat{C})$-plane; (right column): the distribution of the corresponding escape times of the 3D orbits. Top row: $z_0 = 0.01$; Middle row: $z_0 = 0.15$; Bottom row: $z_0 = 0.3$. The color codes are the exactly same as in Figs. \ref{grd1} and \ref{tesc1}.}
\label{xc}
\end{figure*}

The color-coded grids in the physical $(x,y)$ as well as in the phase $(x,\dot{x})$ space provide sufficient information on the phase space mixing for only a fixed value of the Jacobi constant and of course for a specific value of $z_0$. H\'{e}non however, back in the late 60s \citep[e.g.,][]{H69}, introduced a new type of plane which can provide information not only about stability and chaotic regions but also about areas of trapped and escaping orbits using the section $y = \dot{x} = \dot{z} = 0$, $\dot{y} > 0$. In other words, the 3D orbits of the stars of the cluster are launched from the $x$-axis with $x = x_0$, parallel to the $y$-axis $(y = 0)$ having an initial value of $z = z_0$. Consequently, in contrast to the previously discussed types of planes, only orbits with pericenters on the $x$-axis are included and therefore, the value of the dimensionless energy parameter $\widehat{C}$ can be used as an ordinate. In this way, we can monitor how the energy influences the overall orbital structure of the system using a continuous spectrum of energy values rather than few discrete energy levels. In the left column of Fig. \ref{xc} we present the orbital structure of the $(x,\widehat{C})$-plane when $\widehat{C} \in [0.001,0.1]$, for three values of $z_0$, while in the right column of the same figure the distribution of the corresponding escape times of orbits is depicted. The outermost black solid line is the limiting curve which is defined as
\begin{equation}
f_3(x,\widehat{C};z_0) = \Phi_{\rm eff}(x,0,z=z_0) = E.
\label{zvc3}
\end{equation}

We observe that for low energy values close enough to the critical escape energy the vast majority of orbits are trapped chaotic, while as the energy increases the amount of trapped chaotic orbits reduces and the corresponding initial conditions are mainly confined at the boundaries of the stability islands. In addition, we see that there are two main stability islands that are present throughout the energy range however, there are also some smaller islands which correspond to secondary resonant orbits. It should be pointed out that in the blow-ups of the diagrams several additional tiny islands of stability have been identified\footnote{An infinite number of regions of (stable) quasi-periodic (or small scale chaotic) motion is expected from classical chaos theory.}. Table \ref{table1} shows the percentages of the four main types of orbits in the color-coded grids shown in Fig. \ref{xc}. The regions between the stability islands display a high degree of fractalization, while basins of escape are located only at the outer parts of the $(x,\widehat{C})$ planes. The shape as well as the extent of the escape basins depends on the value of $z_0$. In particular, in the cases where orbits are started with low $(z_0 = 0.01)$ or moderate $(z_0 = 0.15)$ initial value of the $z$ coordinate the escape basins are well-defined broad regions, while on the other hand when $z_0 = 0.3$ the basins are thin elongated bands. It is evident that the escape times of the orbits are strongly correlated to the escape basins. Furthermore, one may reasonably conclude that the smallest escape periods correspond to orbits with initial conditions inside the escape basins, while orbits initiated in the fractal regions of the planes have the highest escape rates. In all three cases the escape times are reduced with increases energy.

\begin{table}
\begin{center}
   \caption{Percentages of the four main types of orbits in the color-coded grids shown in Figs. \ref{xc} and \ref{xz}.}
   \label{table1}
   \setlength{\tabcolsep}{4pt}
   \begin{tabular}{@{}lcccc}
      \hline
      Figure & Trapped & Non-escaping & Channel 1 & Channel 2 \\
      \hline
      \ref{xc}-top    &  4.70 & 37.91 & 24.35 & 33.04 \\
      \ref{xc}-middle &  5.61 & 31.60 & 28.98 & 33.81 \\
      \ref{xc}-bottom &  7.34 & 46.57 & 22.83 & 23.26 \\
      \hline
      \ref{xz}-top    & 56.47 & 42.49 &  0.44 &  0.60 \\
      \ref{xz}-middle & 10.27 & 39.98 & 24.53 & 25.22 \\
      \ref{xz}-bottom &  2.14 & 27.85 & 31.45 & 38.56 \\
      \hline
   \end{tabular}
\end{center}
\end{table}

\begin{figure*}
\centering
\resizebox{0.8\hsize}{!}{\includegraphics{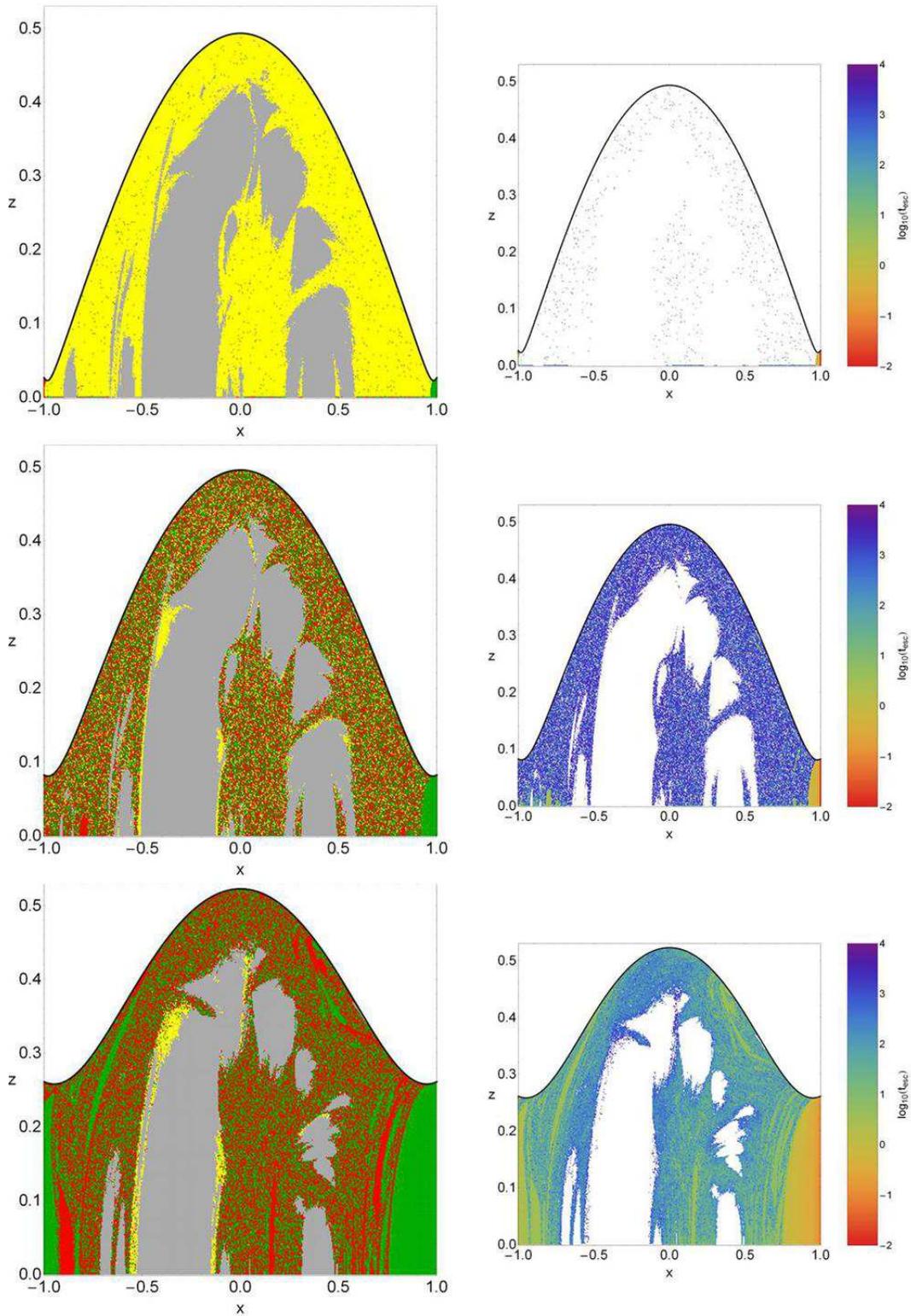}}
\caption{(left column): Orbital structure of the $(x,z)$-plane; (right column): the distribution of the corresponding escape times of the 3D orbits. Top row: $\widehat{C} = 0.001$; Middle row: $\widehat{C} = 0.01$; Bottom row: $\widehat{C} = 0.1$. The color codes are the exactly same as in Figs. \ref{grd1} and \ref{tesc1}.}
\label{xz}
\end{figure*}

It would be very illuminating if we had a more complete view of how the initial value of $z_0$ influences the nature of orbits of the star cluster. In order to obtain this we follow a similar numerical approach to that explained before for the energy thus examining now a continuous spectrum of $z_0$ values. In particular, we use again the section $y = \dot{x} = \dot{z} = 0$, $\dot{y} > 0$, launching orbits once more from the $x$-axis with $x = x_0$, parallel to the $y$-axis, with $z = z_0$. This allow us to construct again a 2D plane in which the $x$ coordinate of orbits is the abscissa, while the $z$ coordinate is the ordinate. In the left column of Fig. \ref{xz} we present the orbital structure of the $(x,z)$-plane when $z \in [0,z_{max}]$ (the 2D case where $z_0 = 0$ is also included), for three values of the energy parameter $\widehat{C}$, while in the right column of the same figure the distribution of the corresponding escape times of orbits is depicted. The maximum allowed value of the $z$ coordinate $z_{max}$ is related with the particular value of the energy level. The outermost black solid line is the limiting curve which this time is given by
\begin{equation}
f_4(x,z;\widehat{C}) = \Phi_{\rm eff}(x,0,z) = E.
\label{zvc4}
\end{equation}

A very complicated orbital structure is reveled in all three studied cases shown in Fig. \ref{xz}. In fact what we see is a dense grid of initial conditions of orbits in the upper part $(x > 0)$ of the the $y = 0$ plane of the 3D equipotential surface (see Fig. \ref{isop}b). We observe the presence of a large number of stability islands (other small and other extended) which occupy significant portion of the planes. For $\widehat{C} = 0.001$ more than half of the $(x,z)$ plane is covered by initial conditions of trapped chaotic orbits. This phenomenon however is expected at such a low energy level just above the critical escape energy. The structure of the stability islands remains almost unperturbed when $\widehat{C} = 0.01$, while on the other hand, the amount of trapped chaotic orbits (about 10\% according to Table \ref{table1}) is heavily decreased and the remaining space displays a high degree of fractalization. At the highest energy level, that is when $\widehat{C} = 0.1$, the areas corresponding to non-escaping regular motion are reduced and extended, well-defined basins of escape emerge. Combining the numerical outcomes presented in Figs. \ref{xc} and \ref{xz} we may conclude that the key factor that determines and controls the escape times of the orbits is the value of the energy (the higher the energy level the shorter the escape rates), while the same quantity (the escape time) seems to be less affected by the initial value of the $z$ coordinate of the orbits.

\section{Trapped chaos}
\label{tc}

\begin{figure*}
\centering
\resizebox{0.8\hsize}{!}{\includegraphics{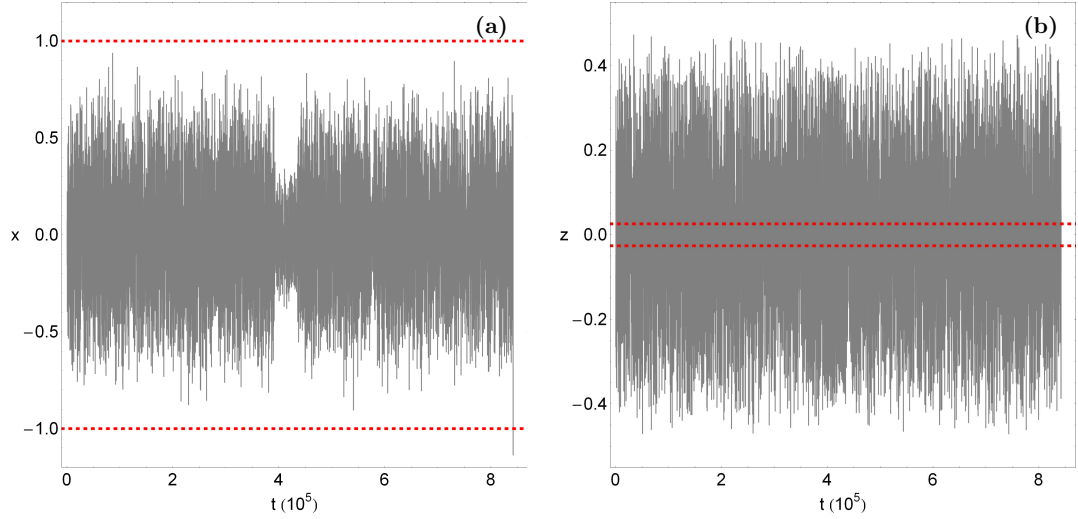}}
\caption{Time-evolution of (a-left): $x$-coordinate and (b-right): $z$-coordinate of a trapped 3D chaotic orbit when $\widehat{C} = 0.001$.}
\label{trap}
\end{figure*}

As it was detailed presented in the previous section, our numerical calculations indicate that at low energy levels, very close to the critical escape energy, the vast majority of the tested initial conditions corresponds to trapped chaotic orbits. Here we will try to interpret and also justify the phenomenon of trapped chaos. By the term ``trapped chaos" we refer to the case where a chaotic orbit remains trapped for a vast time interval inside an open equipotential surface. At this point, it should be emphasized and clarified that these trapped chaotic orbits cannot be considered by no means neither as sticky orbits nor as super sticky orbits with sticky periods larger than $10^4$ time units. Sticky orbits are those who behave regularly for long time periods before their true chaotic nature is fully revealed. In our case on the other hand, this type of orbits exhibit chaoticity very quickly as it takes no more than about 200 time units for the SALI to cross the threshold value (SALI $\ll 10^{-8}$), thus identifying beyond any doubt their chaotic character. Therefore, the main question still remains unanswered: why almost all chaotic orbits are trapped at low energy levels?

\begin{figure*}
\centering
\resizebox{0.8\hsize}{!}{\includegraphics{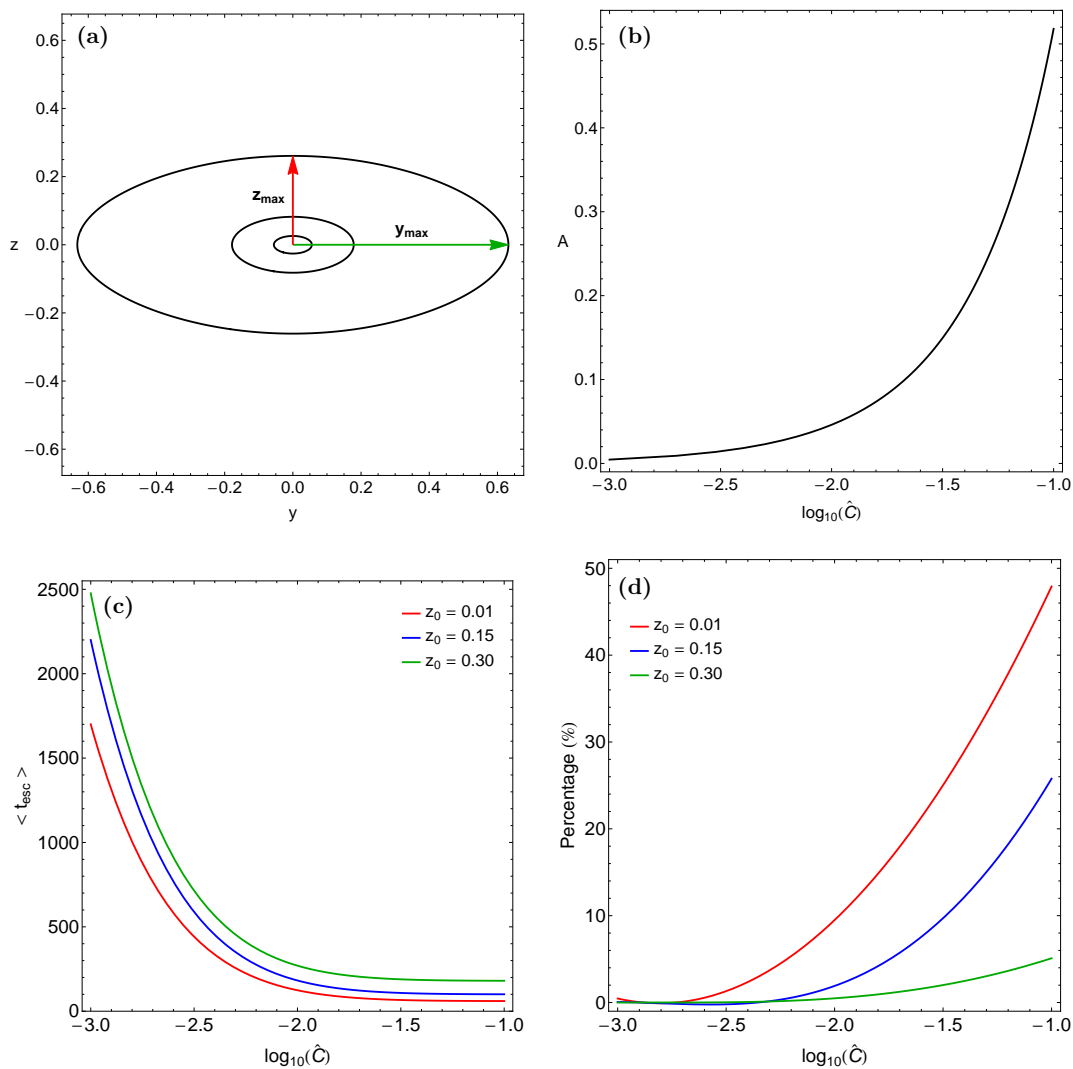}}
\caption{(a-upper left): The structure of the $(y,z)$ contour of 3D equipotential surface when $x = \pm 1$, for three energy levels $\widehat{C} = \{0.001,0.01,0.1\}$ (from inside outwards); (b-upper right): Evolution of (b-upper right): the area of the region on the $(y,z)$ plane; (c-lower left): the average escape time of orbits $< t_{\rm esc} >$ and (d-lower right): the percentage of the phase space covered by the escape basins as a function of the energy parameter $\widehat{C}$.}
\label{cevol}
\end{figure*}

In the following Fig. \ref{trap}(a-b) we observe the time-evolution of the $x$ and $z$ coordinates of a 3D orbit with initial conditions: $x_0 = 0.17$, $y_0 = 0$, $z_0 = 0.01$, $\dot{x_0} = \dot{z_0} = 0$, $\dot{y_0} > 0$ (derived from the Jacobi integral) when $\widehat{C} = 0.001$. The horizontal dashed red lines in panel (a) denote the position of the two Lagrangian points delimiting the transition from trapped to escaping motion, while in panel (b) the same type of lines indicate the minimum and maximum value of $z$ of the equipotential surface at $(x = \pm 1, y = 0)$, or in other words, the width of exit throats. This particular orbit is trapped chaotic and as we can see a huge time of numerical integration, about 842000 time units (more than 84 times the Hubble time) is required so that the orbit can escape from channel 1. We experimented with several trapped chaotic orbits at low energy levels and the obtained results were very similar. Stars moving in chaotic orbits restricted inside the interior region of the cluster having large escape times have also been reported in previous investigations \citep[see e.g., Fig. 11 in][]{FH00}.

\begin{figure*}
\centering
\resizebox{0.8\hsize}{!}{\includegraphics{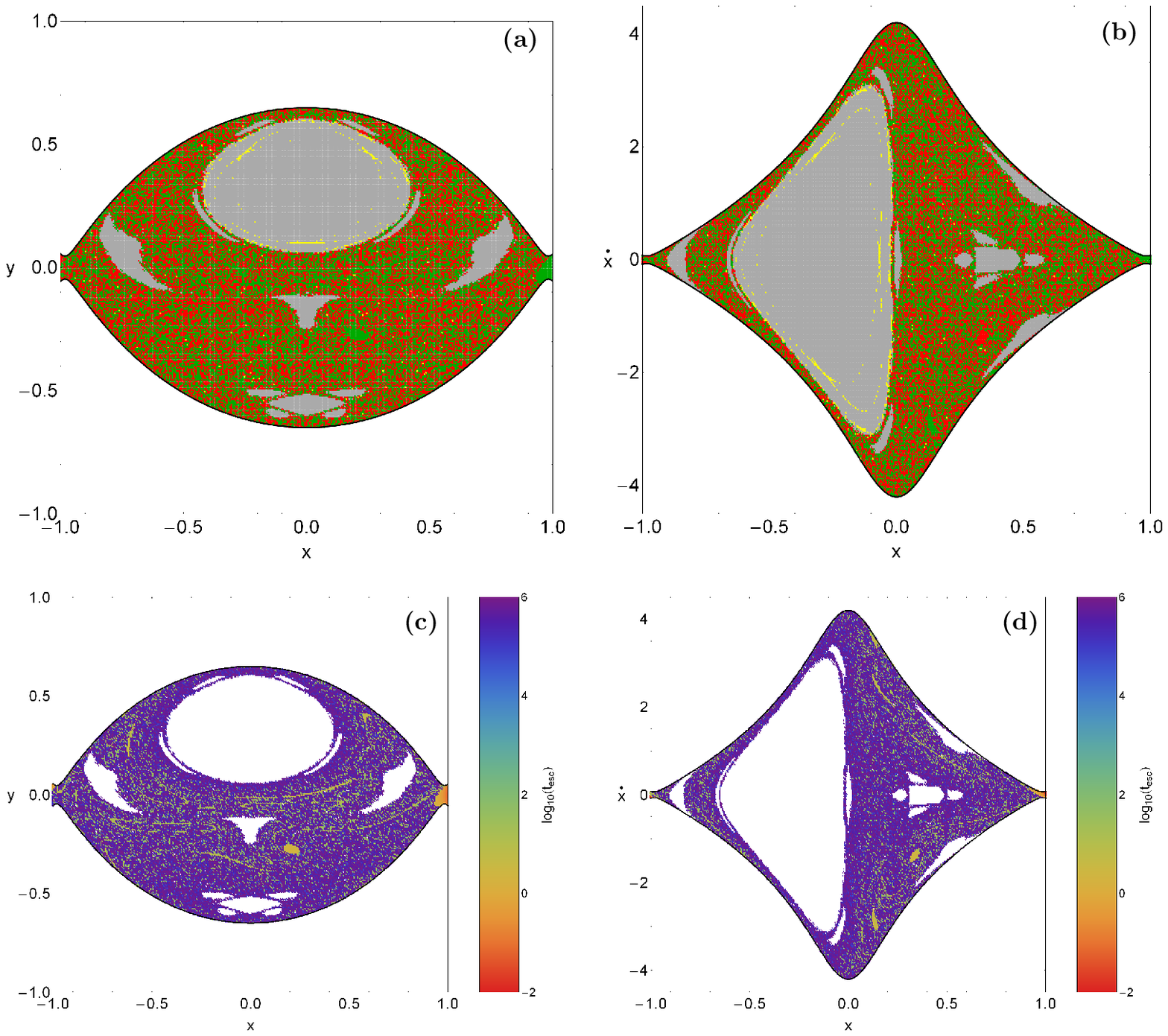}}
\caption{(a-b): Orbital structure of the physical $(x,y)$ and phase $(x,\dot{x})$ space for $\widehat{C} = 0.001$ and $z_0 = 0.01$, when the integration time of the 3D orbits is equal to $10^6$ time units. (c-d): the distribution of the corresponding escape times of orbits. The color codes are the exactly same as in Figs. \ref{grd1} and \ref{tesc1}.}
\label{s0}
\end{figure*}

Our exploration of both the physical and phase space revealed that the escape time of an orbit is directly related to its energy value. It was also found that the value of the energy strongly influences the geometry of the left and right outer parts of the 3D equipotential surface, where the exit channels are located. In particular, the two symmetrical escape channels become more and more wide with increasing energy. This is clearly illustrated in Fig. \ref{cevol}a where we present the structure of the $(y,z)$ contour slice of the 3D equipotential surface when $x = \pm 1$, for three energy levels $\widehat{C} = \{0.001,0.01,0.1\}$ (from inside outwards). It is seen that the width of the escape channels grows rapidly with increasing energy. In order to quantify this increase we decided to calculate the area of the region defined in the $(y,z)$ plane by the escape channel (hole or throat). The region is an ellipse thus the area can be computed as $A = \pi \ y_{max} \ z_{max}$, where the two semi-axes are the maximum values along the axes. In Fig. \ref{cevol}b the evolution of $A$ as a function of the energy parameter $\widehat{C}$ is given. We observe that the area exhibits an exponential increase as we proceed to higher energy levels. In fact, for $\widehat{C} < 0.01$ the area is very limited which means that the width of the exit channels is too small. This geometrical feature justifies the fact why for low energy values, chaotic orbits consume large time periods wandering inside the open equipotential surface until they eventually locate one of the two exits and escape to infinity. Therefore, we may conclude that it is the geometrical shape of the exit channels at low energy levels, which is mainly responsible for the occurrence of the phenomenon of trapped chaos in star clusters.

In Fig. \ref{cevol}c we present the evolution of the average value of the escape time $< t_{\rm esc} >$ of orbits as a function of the dimensionless energy parameter, for three values of $z_0$. One may observes that in all cases the average escape time of orbits reduces rapidly with increasing energy and for $\widehat{C} > 0.01$, the average escape time seems to saturate around 100 time units. We feel it is important to justify this behaviour of the escape time. This can be done by exploiting the geometrical feature of the width of the escape channels discussed earlier in Fig. \ref{cevol}b. In particular, the exponential growth of $A$ explains the rapidly decreasing pattern of $t_{\rm esc}$. Finally Fig. \ref{cevol}d shows the evolution of the percentage of the total area on the phase space corresponding to basins of escape, as a function of the energy parameter for the three initial values of the $z$-coordinate. It is seen, that for high values of $z_0$ $(z_0 = 0.30)$, only at high energy levels we have weak indications of basins of escape corresponding to a tiny fraction of the total phase space, less than 5\%, while the vast majority of the phase space displays a highly fractal structure. When $z_0 = 0.01$ on the other hand, that is the case where the 3D orbits are initiated close to the $(x,y)$ plane, basins of escape appear quite early and they start to grow rapidly occupying about half of the phase space at the highest energy level studied. Thus, we may deduce that for high values of $z_0$ the degree of fractalization is very high, while as we proceed to lower values of $z_0$ the fractility of the phase space is restricted and confined and well-formed basins of escape emerge in the phase space.

Chaotic orbits \textbf{do not} admit a third integral of motion, so according to classical chaos theory for time $t \rightarrow \infty$ they will fill all the available space inside the equipotential surface, unless they escape. In other words, given enough time of numerical integration all trapped chaotic orbits will eventually pass through an exit channel and escape from the cluster. To prove this, we reconstructed the color-coded grids of initial conditions in the physical and in the phase space for $\widehat{C} = 0.001$ and $z_0 = 0.01$ shown previously in Fig. \ref{grd1}(top row) but this time we integrated the 3D orbits for a time interval of $10^6$ time units (roughly about 100 Hubble times). Our results are presented in Fig. \ref{s0}(a-b) where we can see that indeed almost all trapped chaotic orbits have escaped either through $L_1$ or $L_2$ and both types of planes display a high degree of fractalization very similar to that observed in Fig. 5 (bottom row) of Paper I where the 2D case $(z_0 = 0)$ was examined. The distribution of the corresponding escape times of the orbits are given in Fig. \ref{s0}(c-d). We have strong numerical evidence that the same behaviour (trapped chaotic orbits escape when $t_{max} = 10^6$) for $\widehat{C} = 0.001$ is also true when $z_0 = 0.15$ and $z_0 = 0.3$. Looking again Fig. \ref{s0}(a-b), one may identify a very small amount of remaining trapped chaotic orbits (about 0.83\% in the physical and 1.26\% in the phase space) with escape times larger than $10^6$ time units. However, we have to point out that even though that these orbits do eventually escape they must be considered as trapped chaotic ones (not as late escapers), taking into account that these orbits correspond to trajectories of stars moving inside a cluster and therefore, escape times larger than one Hubble time have no physical meaning whatsoever.

\section{Discussion and conclusions}
\label{disc}

The scope of this work was to shed some light to the trapped or escaping nature of three-dimensional orbits in a star cluster embedded in the steady tidal field of a parent galaxy. The Jacobi integral of motion has a critical threshold value (or escape energy) which determines if the escape process is possible or not. In particular, for energies smaller than the threshold value, the equipotential surface is closed and it is certainly true that escape is impossible. For energy levels larger than the escape energy however, the equipotential surface opens and two exit channels appear through which the particles can escape to infinity. Here it should be emphasized, that if a test particle does have energy larger than the escape value, there is no guarantee that the star will certainly escape from the cluster and even if escape does occur, the time required for an orbit to transit through an exit and hence escape to infinity may be vary long compared with the natural escaping time. We managed to distinguish between ordered/chaotic and trapped/escaping orbits and we also located the basins of escape leading to different exit channels, finding correlations with the corresponding escape times of the orbits. Our extensive and thorough numerical investigation strongly suggests, that the overall escape mechanism is a very complicated procedure and very dependent on the value of the Jacobi integral, as well as on the initial value of the $z$ coordinate used for launching the orbits.

Since a distribution function of the model was not available so as to use it for extracting the different samples of orbits, we had to follow an alternative path. We decided to restrict our exploration to a subspace (a 4D grid) of the whole 6D phase space. Thus, we were able to construct again 2D plots depicting the physical $(x,y)$ and the phase $(x,\dot{x})$ plane but with an additional value of $z_0$, since we deal with 3D motion. Following this approach, we identified again regions of order/chaos and bound/escape. We defined for several values of the Jacobi integral, dense uniform grids of initial conditions $(x_0,y_0)$ and $(x_0, \dot{x_0})$ regularly distributed in the area allowed by the energy level on the physical and phase space, respectively. In each type of grid the step separation of the initial conditions along the axes (or in other words the density of the grid) was controlled in such a way that there are always at least $10^5$ orbits to be integrated and classified. For the numerical integration of the orbits in each type of grid, we needed about between 20 minutes and 10 days\footnote{For the two grids presented in Fig. \ref{s0}(a-b), where the orbits were integrated for $10^6$ time units we needed more than 1 month of CPU time for each one.} of CPU time on a Pentium Dual-Core 2.2 GHz PC, depending on the escape rates of orbits in each case. For each initial condition, the maximum time of the numerical integration was set to be equal to $10^4$ time units (at the order of one Hubble time) however, when a particle escaped the numerical integration was effectively ended and proceeded to the next initial condition.

We used the analysis of the 2D system presented in Paper I as a starting point and we expanded our investigation into three dimensions. The novel element of this paper is determination of the influence of the total orbital energy, as well as of the initial value of the $z$ coordinate of the three-dimensional orbits on the escape process. Judging by the detailed outcomes we may say that our task has been successfully completed. The main numerical results of our research can be summarized as follows:
\begin{enumerate}
 \item In all examined cases, areas of bounded motion and regions of initial conditions leading to escape in a given direction (basins of escape), were found to exist in both the physical and the phase space. The several escape basins are very intricately interwoven and they appear either as well-defined broad regions or thin elongated bands. Regions of bounded orbits first and foremost correspond to stability islands of regular orbits where an adelphic integral of motion is present.
 \item At low energy levels just above the critical Jacobi constant we observed the existence of large amounts of chaotic orbits which remain trapped inside the star cluster for time intervals much longer than one Hubble time. This phenomenon was, more or less the same, in all three cases studied regardless of the particular value of $z_0$ of the orbits. Thus, we employed the SALI method in order to distinguish between non-escaping regular orbits and trapped chaotic ones.
 \item A strong correlation between the extent of the basins of escape and the value of the Jacobi integral was found to exist. Indeed, for low energy levels the structure of both the physical and the phase space exhibits a large degree of fractalization and therefore the majority of orbits escape choosing randomly escape channels. As the value of the energy increases however, the structure becomes less and less fractal and several basins of escape emerge. The extent of these basins of escape is more prominent at low values of $z_0$ and at high energy levels.
 \item We observed, that in many cases the escape process is highly sensitive dependent on the initial conditions, which means that a minor change in the initial conditions of an orbit lead the test particle to escape through the other exit channel. These regions are the exact opposite of the escape basins, are completely intertwined with respect to each other (fractal structure) and are mainly located in the vicinity of stability islands. This sensitivity towards slight changes in the initial conditions in the fractal regions implies, that it is impossible to predict through which exit the particle will escape.
 \item It was detected, that for energy levels slightly above the escape energy the majority of the escaping orbits have considerable long escape rates (or escape periods), while as we proceed to higher energies the proportion of fast escaping orbits increases significantly. This behavior can be justified, if we take into account that with increasing energy the exit channels on the equipotential surface become more and more wide thus the test particles can find easily and faster one of the two exits and escape to infinity.
  \item Our calculations revealed, that the escape times of orbits are directly linked to the basins of escape. In particular, inside the basins of escape as well as relatively away from the fractal domains, the shortest escape rates of the orbits had been measured. On the other hand, the longest escape periods correspond to initial conditions of orbits either near the boundaries between the escape basins or in the vicinity of stability islands.
 \item The phenomenon of trapped chaotic orbits at low energy levels was further explored and this behavior was justified by a geometrical feature of the equipotential surface. Specifically, for low values of energy the width of the exit channels is too small therefore, chaotic orbits consume large time periods wandering inside the open equipotential surface until they eventually locate one of the two exits and escape to infinity. Moreover, it was proved that the width of the escape channels grows rapidly with increasing energy.
\end{enumerate}

We hope that the present numerical analysis and the corresponding results to be useful in the active field of the dissolution of tidally limited star clusters. The outcomes as well as the conclusions of the present research are considered, as an initial effort and also as a promising step in the task of understanding the escape mechanism of three-dimensional orbits in open star clusters. Taking into account that our results are encouraging, it is in our future plans to focus our attention on exploring in detail the fate of escaping stars moving into the so-called tidal tails or tidal arms.

\section*{Acknowledgments}

I would like to express my warmest thanks to Dr. Andreas Ernst and Prof. Douglas C. Heggie for all the illuminating and inspiring discussions during this research. My thanks also go to the anonymous referee for the careful reading of the manuscript and for all the apt suggestions and comments which allowed us to improve both the quality and the clarity of the paper.

\bsp
\label{lastpage}


\begin{thebibliography}{}

\bibitem[\protect\citeauthoryear{Aarseth}{1999}]{A99} Aarseth S.J., 1999, PASP, 111, 1333

\bibitem[\protect\citeauthoryear{Aarseth}{2003}]{A03} Aarseth S.J., 2003, Gravitational N-Body Simulations

\bibitem[\protect\citeauthoryear{Aguirre et al.}{2001}]{AVS01} Aguirre J., Vallejo J.C., Sanju\'{a}n M.A.F, 2001, Phys. Rev E, 64, 066208

\bibitem[\protect\citeauthoryear{Aguirre \& Sanju\'{a}n}{2003}]{AS03} Aguirre J., Sanju\'{a}n M.A.F., 2003, Phys. Rev. E, 67, 056201

\bibitem[\protect\citeauthoryear{Aguirre et al.}{2009}]{AVS09} Aguirre J., Viana R.L., Sanju\'{a}n M.A.F., 2009, Rev. Mod. Phys., 81, 333

\bibitem[\protect\citeauthoryear{Ambartsumian}{1938}]{A38} Ambartsumian V.A., 1938, Ann. Leningrad State Univ., 22, 19

\bibitem[\protect\citeauthoryear{Barrio et al.}{2008}]{BBS08} Barrio R., Blesa F., Serrano S., 2008, Europhys. Lett., 82, 10003

\bibitem[\protect\citeauthoryear{Baumgardt}{2001}]{B01} Baumgardt H., 2001, MNRAS, 325, 1323

\bibitem[\protect\citeauthoryear{Belokurov et al.}{2006}]{BEI06} Belokurov V., Evans N.W., Irwin M.J., Hewett P.C., Wilkinson M.I., 2006, ApJ, 637, L29

\bibitem[\protect\citeauthoryear{Benet et al.}{1996}]{BTS96} Benet L., Trautmann D., Seligman T.H., 1996, CeMDA, 66, 203

\bibitem[\protect\citeauthoryear{Benet et al.}{1998}]{BST98} Benet L., Seligman T., Trautman D., 1998, CeMDA, 71, 167

\bibitem[\protect\citeauthoryear{Bleher et al.}{1988}]{BGOB88} Bleher S., Grebogi C., Ott E., Brown R., 1988, Phys. Rev. A, 38, 930

\bibitem[\protect\citeauthoryear{Blesa et al.}{2012}]{BSBS12} Blesa F., Seoane J.M., Barrio R., Sanju\'{a}n M.A.F. 2012, Int. J. Bifurc. Chaos, 22, 1230010

\bibitem[\protect\citeauthoryear{Binney \& Tremaine}{2008}]{BT08} Binney J., Tremaine S., 2008, Galactic Dynamics, Princeton Univ. Press, Princeton, USA

\bibitem[\protect\citeauthoryear{Bressert et al.}{2010}]{BBG10} Bressert E., Bastian N., Gutermuth R., Megeath S.T., Allen L., et al., 2010, MNRAS, 409, L54

\bibitem[\protect\citeauthoryear{Capuzzo Dolcetta et al.}{2005}]{CMM05} Capuzzo Dolcetta R., Di Matteo P., Miocchi P., 2005, AJ, 129, 1906

\bibitem[\protect\citeauthoryear{Carpintero et al.}{2014}]{CMD14} Carpintero D.D., Maffione N., Darriba L., 2014, Astronomy and Computing, 5, 19

\bibitem[\protect\citeauthoryear{Carpintero \& Aguilar}{1998}]{CA98} Carpintero D.D., Aguilar L.A. 1998, MNRAS, 298, 1

\bibitem[\protect\citeauthoryear{Chandrasekhar}{1942}]{C42} Chandrasekhar S., 1942, Principles of Stellar Dynamics. Univ. Chicago Press, Chicago

\bibitem[\protect\citeauthoryear{Choi et al.}{2007}]{CWK07} Choi J.-H., Weinberg M.D., Katz N., 2007, MNRAS, 381, 987

\bibitem[\protect\citeauthoryear{Churchill et al.}{1979}]{C79} Churchill R.C., et al. 1979, in Como Conference Proceedings on Stochastic Behavior in Classical and Quantum Hamiltonian Systems, Volume 93, Lecture Notes in Physics, ed. G. Casati, J. Fords (Berlin: Springer)

\bibitem[\protect\citeauthoryear{Contopoulos}{1990}]{C90} Contopoulos G., 1990, A\&A, 231, 41

\bibitem[\protect\citeauthoryear{Contopoulos \& Kaufmann}{1992}]{CK92} Contopoulos G., Kaufmann D., 1992, A\&A, 253, 379

\bibitem[\protect\citeauthoryear{Dehnen et al.}{2004}]{DOG04} Dehnen W., Odenkirchen M., Grebel E.K., Rix H.-W., 2004, AJ, 127, 2753

\bibitem[\protect\citeauthoryear{de Moura \& Letelier}{2000}]{dML00} de Moura A.P.S., Letelier P.S., 2000, Phys. Rev. E, 62, 4784

\bibitem[\protect\citeauthoryear{di Mateo et al.}{2005}]{dMCM05} di Matteo P., Capuzzo Dolcetta R., Miocchi P., 2005, Cel. Mech. Dyn. Astron., 91, 59

\bibitem[\protect\citeauthoryear{Ernst et al.}{2007}]{EGFJ07} Ernst A., Glaschke P., Fiestas J., Just A., Spurzem R., 2007, MNRAS, 377, 465

\bibitem[\protect\citeauthoryear{Ernst et al.}{2008}]{EJSP08} Ernst A., Just A., Spurzem R., Porth O., 2008, MNRAS, 383, 897 (Paper I)

\bibitem[\protect\citeauthoryear{Feast \& Whitelock}{1997}]{FW97} Feast M., Whitelock P., 1997, MNRAS, 291, 683

\bibitem[\protect\citeauthoryear{Fellhauer et al.}{2007}]{FEB07} Fellhauer M., Evans N.W., Belokurov V., Wilkinson M.I., Gilmore G., 2007, MNRAS, 380, 749

\bibitem[\protect\citeauthoryear{Fukushige \& Heggie}{2000}]{FH00} Fukushige T., Heggie D.C., 2000, MNRAS, 318, 753

\bibitem[\protect\citeauthoryear{Grillmair et al.}{1995}]{GFI95} Grillmair C.J., Freeman K.C., Irwin M., Quinn P.J., 1995, AJ, 109, 2553

\bibitem[\protect\citeauthoryear{Guido et al.}{2013}]{GLH13} Guido R.I., Loyola M., Hurley J.R., 2013, MNARS, 434, 2509

\bibitem[\protect\citeauthoryear{H\'{e}non}{1960}]{H60} H\'{e}non M., 1960, Ann. Astrophys., 23, 668

\bibitem[\protect\citeauthoryear{H\'{e}non}{1969}]{H69} H\'{e}non M., 1969, A\&A, 1, 223

\bibitem[\protect\citeauthoryear{Holmberg \& Flynn}{2000}]{HF00} Holmberg J., Flynn C., 2000, MNRAS, 313, 209

\bibitem[\protect\citeauthoryear{Hurley et al.}{2005}]{HPA05} Hurley J.R., Pols O.R., Aarseth S.J., Tout C.A., 2005, MNRAS, 363, 293

\bibitem[\protect\citeauthoryear{Hut \& Bahcall}{1983}]{HB83} Hut P., Bahcall J.N., 1983, ApJ, 268, 319

\bibitem[\protect\citeauthoryear{Innanen}{1980}]{I80} Innanen K.A., 1980, ApJ, 85, 81

\bibitem[\protect\citeauthoryear{Johnston et al.}{1999}]{JSH99} Johnston K.V., Sigurdsson S., Hernquist L., 1999, MNRAS, 302, 771

\bibitem[\protect\citeauthoryear{Jung \& Scholz}{1988}]{JS88} Jung C., Scholz H., 1988, J. Phys. A, 21, 3607

\bibitem[\protect\citeauthoryear{Just et al.}{2009}]{JBPE09} Just A., Berczik P., Petrov M., Ernst A., 2009, MNRAS, 392, 969

\bibitem[\protect\citeauthoryear{Kharchenko et al.}{1997}]{KSL97} Kharchenko N., Scholz R.-D., Lehmann I., 1997, A\&AS, 121, 439

\bibitem[\protect\citeauthoryear{King}{1959}]{K59} King I.R., 1959, AJ, 64, 351

\bibitem[\protect\citeauthoryear{King}{1962}]{K62} King I.R., 1962, AJ, 67, 471

\bibitem[\protect\citeauthoryear{Koch et al.}{2004}]{KGO04} Koch A., Grebel E.K., Odenkirchen M., Mart\'{i}nez-Delgado D., Caldwell J.A.R., 2004, AJ, 128, 2274

\bibitem[\protect\citeauthoryear{Kollmeier \& Gould}{2007}]{KG07} Kollmeier J.A., Gould A., 2007, ApJ, 664, 343

\bibitem[\protect\citeauthoryear{Kroupa et al.}{1993}]{KTG93} Kroupa P., Tout C.A., Gilmore G., 1993, MNRAS, 262, 545

\bibitem[\protect\citeauthoryear{Kruijssen}{2012}]{K12} Kruijssen J.M.D., 2012, MNRAS, 426, 3008

\bibitem[\protect\citeauthoryear{K\"{u}pper et al.}{2008}]{KMH08} K\"{u}pper A.H.W., Macleod A., Heggie D.C., 2008, MNRAS, 387, 1248

\bibitem[\protect\citeauthoryear{K\"{u}pper et al.}{2010}]{KKBH10} K\"{u}pper A.H.W., Kroupa P., Baumgardt H., Heggie D.C., 2010, MNRAS, 401, 105

\bibitem[\protect\citeauthoryear{Lada \& Lada}{2003}]{LL03} Lada C.J., Lada E.A., 2003, ARA\&A, 41, 57

\bibitem[\protect\citeauthoryear{Lee et al.}{2003}]{LLF03} Lee K.H., Lee H.M., Fahlman G.G., Lee M.G., 2003, AJ, 126, 815

\bibitem[\protect\citeauthoryear{Lee et al.}{2006}]{LLS06} Lee K.H., Lee H.M., Sung H., 2006, MNRAS, 367, 646

\bibitem[\protect\citeauthoryear{Lehmann \& Scholz}{1997}]{LS97} Lehmann I., Scholz R.-D., 1997, A\&A, 320, 776

\bibitem[\protect\citeauthoryear{Li et al.}{2012}]{LLZ12} Li Y., Luo A., Zhao G., Lu Y., Ren J., Zuo F., 2012, ApJ, 744, L24

\bibitem[\protect\citeauthoryear{Meylan \& Heggie}{1997}]{MH97} Meylan G., Heggie D.C., 1997, A\&A Rev., 8, 1

\bibitem[\protect\citeauthoryear{Montuori et al.}{2007}]{MCd07} Montuori M., Capuzzo-Dolcetta R., di Matteo P., Lepinette A., Miocchi P., 2007, ApJ, 659, 1212

\bibitem[\protect\citeauthoryear{Motter \& Lai}{2002}]{ML02} Motter A.E., Lai Y.C.: 2002, Phys. Rev. E, 65 R015205

\bibitem[\protect\citeauthoryear{Nagler}{2004}]{N04} Nagler J., 2004, Physical Review E, 69, 066218

\bibitem[\protect\citeauthoryear{Nagler}{2005}]{N05} Nagler J., 2005, Physical Review E, 71, 026227

\bibitem[\protect\citeauthoryear{Odenkirchen et al.}{2001}]{OGR01} Odenkirchen M., Grebel, E.K., Rockosi, C.M., Dehnen, W., Ibata, R., et al., 2001, ApJ, 548, L165

\bibitem[\protect\citeauthoryear{Press et al.}{1992}]{PTVF92} Press H.P., Teukolsky S.A, Vetterling W.T., Flannery B.P., 1992, Numerical Recipes in FORTRAN 77, 2nd Ed., Cambridge Univ. Press, Cambridge, USA

\bibitem[\protect\citeauthoryear{Rockosi et al.}{2002}]{ROG02} Rockosi C.M., Odenkirchen, M., Grebel, E.K., Dehnen, W., Cudworth, K.M., Gunn, J.E., et al., 2002, AJ, 124, 349

\bibitem[\protect\citeauthoryear{Ross et al.}{1997}]{RMH97} Ross D.J., Mennim A., Heggie D.C., 1997, MNRAS, 284, 811

\bibitem[\protect\citeauthoryear{Royer}{1997}]{R97} Royer F., 1997, in R.M. Bonnet, E. H{\o}g, P.L. Bernacca, L. Emiliani, A. Blaauw, C. Turon, J. Kovalevsky, L. Lindegren, H. Hassan, M. Bouffard, B. Strim, D. Heger, M.A.C. Perryman \& L. Woltjer ed., Hipparcos - Venice '97 Vol. 402 of ESA Special Publication, Populations among High-Velocity Early-Type Stars. pp 595-598

\bibitem[\protect\citeauthoryear{Seoane et al.}{2006}]{SASL06} Seoane J.M., Aguirre J., Sanju\'{a}n M.A.F., Lai Y.C., 2006, Chaos, 16, 023101

\bibitem[\protect\citeauthoryear{Seoane et al.}{2007}]{SSL07} Seoane J.M., Sanju\'{a}n M.A.F., Lai Y.C., 2007, Phys. Rev. E, 76, 016208

\bibitem[\protect\citeauthoryear{Seoane \& Sanju\'{a}n}{2008}]{SS08} Seoane J.M., Sanju\'{a}n M.A.F., 2008, Phys. Lett. A, 372, 110

\bibitem[\protect\citeauthoryear{Seoane et al.}{2009}]{SHSL09} Seoane J.M., Huang L., Sanju\'{a}n M.A.F., Lai Y.C., 2009, Phys. Rev. E, 79, 047202

\bibitem[\protect\citeauthoryear{Seoane \& Sanju\'{a}n}{2010}]{SS10} Seoane J.M., Sanju\'{a}n M.A.F., 2010, Int. J. Bifurc. Chaos, 9, 2783

\bibitem[\protect\citeauthoryear{Silva \&  Napiwotzki}{2011}]{SN11} Silva M.D.V., Napiwotzki R., 2011, MNRAS, 411, 2596

\bibitem[\protect\citeauthoryear{Skokos}{2001}]{S01} Skokos C., 2001, J. Phys. A: Math. Gen., 34, 10029

\bibitem[\protect\citeauthoryear{Spitzer}{1940}]{S40} Spitzer L., 1940, MNRAS, 100, 396

\bibitem[\protect\citeauthoryear{Tout et al.}{1997}]{TAP97} Tout C.A., Aarseth S.J., Pols O.R., Eggleton P.P., 1997, MNRAS, 291, 732

\bibitem[\protect\citeauthoryear{Yim \& Lee}{2002}]{YL02} Yim K.-J., Lee H.M., 2002, JKAS, 35, 75

\bibitem[\protect\citeauthoryear{Zotos}{2012}]{Z12} Zotos E.E., 2012, PASA, 29, 161

\bibitem[\protect\citeauthoryear{Zotos}{2014}]{Z14} Zotos E.E., 2014, Nonlinear Dynamics, 76, 1301

\bibitem[\protect\citeauthoryear{Zotos \& Caranicolas}{2013}]{ZCar13} Zotos E.E., Caranicolas N.D., 2013, Nonlinear Dynamics, 74, 1203

\bibitem[\protect\citeauthoryear{Zotos \& Carpintero}{2013}]{ZC13} Zotos E.E., Carpintero D.D., 2013, CeMDA, 116, 417

\end{thebibliography}
\end{document}